\documentclass[12pt]{article}
\usepackage{latexsym}
\usepackage{float}
\usepackage{amsmath,bm,amssymb}
\usepackage{colordvi}
\usepackage{multicol,multirow}
\usepackage{color}
\usepackage{rotating}
 \usepackage{url,hyperref}
\usepackage[ruled]{algorithm2e}
\usepackage{pdflscape}
\usepackage{subfigure}
\usepackage{graphicx}
\usepackage[utf8]{inputenc}
\usepackage{longtable}

\def\doublespace{\baselineskip=22pt}
\usepackage{lscape}

\begin{document}
\doublespace
\baselineskip 2.8ex
\begin{center}
{\bf\LARGE Identifying Gene-environment interactions with robust marginal Bayesian variable selection }
\doublespace

\vspace{0.8em}


{\bf\large Xi Lu$^1$, Kun Fan$^1$, Jie Ren$^2$ and Cen Wu$^{1\ast}$}

\vspace{0.6em}

{ $^1$ Department of Statistics, Kansas State University, Manhattan, KS}\\

{ $^2$  Department of Biostatistics, Indiana University School of Medicine, Indianapolis, IN}\\

\end{center}

{\bf $\ast$ Corresponding author}:
Cen Wu, wucen@ksu.edu\\
\vspace{0.8em}

{\bf\Large Abstract}\\

In high-throughput genetics studies, an important aim is to identify gene-environment interactions associated with the clinical outcomes. Recently, multiple marginal penalization methods have been developed and shown to be effective in G$\times$E studies. However, within the Bayesian framework, marginal variable selection has not received much attention. In this study, we propose a novel marginal Bayesian variable selection method for G$\times$E studies. In particular, our marginal Bayesian method is robust to data contamination and outliers in the outcome variables. With the incorporation of spike-and-slab priors, we have implemented the Gibbs sampler based on MCMC. The proposed method outperforms a number of alternatives in extensive simulation studies. The utility of the marginal robust Bayesian variable selection method has been further demonstrated in the case studies using data from the Nurse Health Study (NHS). Some of the identified main and interaction effects from the real data analysis have important biological implications. \\

\noindent{\bf Keywords:} Gene-environment interaction; marginal analysis; MCMC; robust Bayesian variable selection; spike-and-slab priors.

\section{Introduction}

The risk and progression of complex diseases including cancer, asthma and type 2 diabetes, are associated with the coordinated functioning of genetic factors, the environmental (and clinical) factors, as well as their interactions(\cite{HUNTER,SIM,VON, CORN}). The identification of important Gene-environment(G$\times$E) interactions leads to novel insight in dissecting the genetic basis of complex diseases in addition to the main effects of genetic and environmental factors. In the last two decades, searching for the important G$\times$E interactions has been extensively conducted based on genetic association studies(\cite{CORD,WLC}). One representative example is the genome wide association study (GWAS),  where the statistical significance of interaction between the environmental exposure and the genetic variant has been marginally assessed one at a time across the whole genome. Important findings are evidenced by genome wide significant p-values after adjusting for multiple comparisons.   \par

Recently, substantial efforts have been devoted to novel penalized variable selection methods for G$\times$E studies(\cite{ZFR}). In particular, marginal penalization has achieved very competitive performances with the aforementioned significance based G$\times$E analysis(\cite{SLH, CZJ,ZSX}). For example, within the framework of maximum rank correlation, Shi et al.(2014)(\cite{SLH}) has developed a penalization method robust to outliers and model misspecification in determining important G$\times$E interactions one at a time. Zhang et al. (2019)(\cite{ZSX}) has imposed hierarchical structure between the main effects and interactions in marginal identification of G$\times$E interactions using regularization. Despite success, these studies have limitations. First, as a common tuning parameter is demanded for all the marginal models, its selection requires pooling all genes together to conduct a joint model based cross validation. While such a strategy is not rare, it seems not in favor of the marginal nature of the proposed G$\times$E studies.  Second, a rigorous measure to quantify uncertainty is not available. Zhang et al. (2019)(\cite{ZSX}) has constructed 95\% confidence intervals based on the observed occurrence index (OOI) values(\cite{HJM}), nevertheless, this measure has been used to demonstrate stability of identified effects rather than quantifying uncertainty of penalized estimates.    \par

These limitations have motivated us to consider Bayesian analyses. In literature, Bayesian variable selection methods have been developed for G$\times$E analysis in multiple studies(\cite{ZFR}). For example, with indicator model selection, Liu et al. (2015) (\cite{LMA}) has imposed hierarchical Bayesian variable selection for linear G$\times$E interactions.  Li et al. (2015)(\cite{LWL}) has proposed a Bayesian group LASSO to identify non-linear interactions in nonparametric varying coefficient models. Ren et al. (2020)(\cite{RZL}) has further incorporated selection of linear and nonlinear G$\times$E interactions simultaneously while accounting for structured identification in the Bayesian adaptive shrinkage framework.  All these fully Bayesian methods can efficiently provide uncertainty quantification based on the posterior samples from MCMC. Nevertheless, our limited literature mining shows that none of the marginal Bayesian variable selection methods have been proposed for interaction studies so far. 

Historically, marginal analysis has prevailed in G$\times$E interaction studies within the framework of genetic association studies. Although recent studies have confirmed the utility of regularized variable selection in joint G$\times$E analysis, more efforts are needed for marginal penalizations especially through the Bayesian point of view. The step towards marginal Bayesian variable selection is of particular significance in developing a coherent framework of analyzing G$\times$E interactions.

Here, we propose a novel marginal Bayesian variable selection method for the robust identification of G$\times$E interactions. As heavy-tailed distributions and outliers in the response variable have been widely observed, robust modelling is essential for yielding reliable results. Specifically, the robustness of the proposed method is facilitated by the Bayesian formulation of the least absolute deviation (LAD) regression which has been a popular choice in frequentist G$\times$E studies but seldom investigated in a similar context from the Bayesian perspective. We consider the Bayesian LAD LASSO for regularized identification of interaction effects. As Bayesian LAD LASSO does not lead to zero coefficients, the spike-and-slab priors(\cite{GEMC, ISHW}) has been incorporated to impose exact sparsity in the adaptive shrinkage framework. The corresponding MCMC algorithm has been developed to accommodate fast computations. We have demonstrated the advantage of the proposed robust Bayesian marginal analysis in simulation. The findings from the case study of  the Nurses’ Health Study (NHS) with SNP measurements have important biological implications.

\section{Method}

We use $Y$ to denote a continuous response variable representing the the cancer outcome or disease phenotype. Let $X=(X_{1},\ldots,X_{p})$ be the $p$ genetic variants, $E=(E_{1},\ldots,E_{q})$ be the $q$ environmental factors and $C=(C_{1},\ldots,C_{m})$ be the $m$ clinical factors. We denote the $i$th subject with $i$. Let ($Y_{i}$, $E_{i}$, $C_{i}$, $X_{i}$) ($i = 1,\ldots,n$) be independent and identically distributed random vectors. For $X_{ij}$ ($j = 1,\ldots,p$), the measurement of the $j$th genetic factor on the $i$th subject,  consider the following marginal model:
\begin{equation}\label{equr:dataModel}
\begin{split}
Y_{i}&=\sum_{k=1}^{q}\alpha_{k}E_{ik}+\sum_{t=1}^{m}\gamma_{t}C_{it}+\beta_{j}X_{ij}+\sum_{k=1}^{q}\eta_{jk}X_{ij}E_{ik}+\epsilon_{i}\\
&=\sum_{k=1}^{q}\alpha_{k}E_{ik}+\sum_{t=1}^{m}\gamma_{t}C_{it}+\beta_{j}X_{ij}+\eta_{j}\tilde{W}_{i}+\epsilon_{i},
\end{split}
\end{equation}
where $\alpha_{k}$'s and $\gamma_{t}$'s are the regression coefficients corresponding to effects of environmental and clinical factors, respectively. For the $j$th gene $X_{j}$ ($j = 1,\ldots,p$), the G$\times$E interactions effects are defined with $W_{j} = (X_{j}E_{1},\ldots,X_{j}E_{q})$, $\eta_{j} = (\eta_{j1}, \ldots, \eta_{jq})^{T}$. With a slight abuse of notation, denote $\tilde{W} = W_{j}$. The $\beta_{j}$'s and $\eta_{jk}$'s are the regression coefficients of the genetic variants and G$\times$E interactions effects, correspondingly.  Denote $\alpha=(\alpha_{1}, \ldots, \alpha_{q})^{T}$ and $\gamma=(\gamma_{1}, \ldots, \gamma_{m})^{T}$. 
Then model (\ref{equr:dataModel}) can be written as 
\begin{equation}
Y_{i} = E_{i}\alpha + C_{i}\gamma + X_{ij}\beta_{j} + \tilde{W}_{i}\eta_{j} + \epsilon_{i}.
\end{equation}

\subsection{Bayesian formulation of the LAD regerssion}

The necessity of accounting for robustness in interaction studies has been increasingly recognized (\cite{ZFR}). Within the frequentist framework, it is essentially dependent on adopting a robust loss function to quantify lack of fit (\cite{WUM}). Among a variety of popular robust losses, the least absolute deviation (LAD) loss function is well known for its advantages in dealing with heavy-tailed error distributions or outliers in response. The estimation of regression coefficients amounts to the following minimization problem
\begin{equation*}
\min_{\alpha, \gamma, \beta_{j}, \eta_{j}} \sum_{i=1}^{n}|Y_{i}-E_{i}\alpha - C_{i}\gamma - X_{ij}\beta_{j} - \tilde{W}_{i}\eta_{j}|.
\end{equation*}

Here, we propose the robust marginal Bayesian variable selection based on LAD.  As the Laplace distribution is equivalent to the mixture of an exponential distribution and a scaled normal distribution(\cite{KOZU}), for a Bayesian formulation of LAD regression, we assume that $\epsilon_{i} (i=1,\dots,n)$ are i.i.d. random variables following the Laplace distribution with density

\begin{equation*}
f(\epsilon_{i}|\tau) = \frac{\tau}{2} \text{exp} (-\tau|\epsilon_{i}|),
\end{equation*}
where $\tau$ is the inverse of the scale parameters from the Laplace density. Then the likelihood function of our marginal G$\times$E model can be expressed as:
\begin{equation*}
f(Y|\alpha, \gamma, \beta_{j}, \eta_{j}) = \prod_{i=1}^{n} \frac{\tau}{2} \text{exp} (-\tau|Y_{i}-E_{i}\alpha - C_{i}\gamma - X_{ij}\beta_{j} - \tilde{W}_{i}\eta_{j}|).
\end{equation*}
The above formulation using Laplace distrubution is a special case of the asymmetric Laplace distribution, which has been widely adopted in Baysian quantile regression(\cite{MOY,YUK}). In Baysian quantile regression, $\epsilon_{i}$'s are assumed to follow the skewed Laplace distribution with density
\begin{equation*}\label{equr:full}
f(\epsilon|\tau) = \theta(1-\theta)\tau \text{exp}(-\tau \rho_{\theta}(\epsilon)).
\end{equation*}
The random errors can be written as 
\begin{equation*}
\epsilon_{i} = \xi_{1} v_{i} + \tau^{-1/2}\xi_{2}\sqrt{v_{i}}z_{i},
\end{equation*}
where
\begin{equation*}\label{equr:full}
\xi_{1}  = \frac{1 - 2\theta}{\theta(1-\theta)} \quad and \quad  \xi_{2} = \sqrt{\frac{2}{\theta(1-\theta)}}
\end{equation*}
with quantile level $\theta\in(0,1)$, $v_{i}{\thicksim}\text{exp}(\tau^{-1})$, and $z_{i}{\thicksim}\text{N}(0,1)$.\par
The Bayesian LAD regression is a special case of Bayesian quantile regression(\cite{NANLIN}) with $\theta$=0.5, resulting in that $\xi_{1} =0$ and $\xi_{2} =\sqrt{8}$. Therefore, the response $Y_{i}$ can be written as:
\begin{equation}\label{equr:ge2}
\begin{aligned}
Y_{i} &= \mu_{i} + \tau^{-1/2}\xi_{2}\sqrt{v_{i}}z_{i}, \\
v_{i}|\tau & \stackrel{iid}{\thicksim}\tau \text{exp}(- \tau v_{i}),\\
z_{i} & \stackrel{iid}{\thicksim}\text{N}(0,1),
\end{aligned}
\end{equation}
where $\mu_{i} = E_{i}\alpha + C_{i}\gamma + X_{ij}\beta_{j} + \tilde{W}_{i}\eta_{j}$.

\subsection{Bayesian LAD LASSO with spike-and-slab priors}

In model (\ref{equr:dataModel}), the coefficients $\beta_{j}$ and $\eta_{j}$ corresponds to the main and interaction effects with respect to the $j$th genentic variant, respectively. When $\beta_{j} = 0$ and $\eta_{j} = 0$,  the genetic variant has no effect on the phenotype. A non-zero $\beta_{j}$ suggests the presence of main genetic effect. For $\eta_{j}$, if at least one of its component is not zero, then the G$\times$E interaction effect exist. In the literature, Bayesian quantile LASSO, with Bayesian LAD LASSO as its special case, has been proposed to conduct variable selection(\cite{NANLIN}). However, a major limitation is that Bayesian quantile LASSO cannot shrink regression coefficients to 0 exactly, resulting in inaccurate identification and biased estimation. To overcome such an limitation, we incorporate spike-and-slab priors to impose sparsity within Bayesian LAD LASSO framework as follows. 

For the $j$th gene ($j=1,\dots,p$), the marginal LAD LASSO model is given by: 
\begin{equation*}
 \sum_{i=1}^{n}|Y_{i}-E_{i}\alpha - C_{i}\gamma - X_{ij}\beta_{j} - \tilde{W}_{i}\eta_{j}|+\lambda_1|\beta_{j}|+\lambda_2 \sum_{k=1}^{q}|\eta_{jk}|.
\end{equation*}
Let $\varphi_{1}=\tau\lambda_1$ and $\varphi_{2}=\tau\lambda_2$. Then the conditional Laplace prior on the coefficient of main effect $\beta_{j}$ can be expressed as scale mixtures of normals:
\begin{equation*}
\begin{aligned}
\pi(\beta_{j}|\tau, \lambda_1) &= \frac{\varphi_1}{2}\text{exp}\{-\varphi_1|\beta_{j}|\}\\
&=\int_0^{\infty}\frac{1}{\sqrt{2\pi s_1}}\text{exp}(-\frac{\beta_{j}^2}{2 s_1})\frac{\varphi_1^2}{2}\text{exp}(\frac{-\varphi_1^2}{2}s_1)d s_1.
\end{aligned}
\end{equation*}
 The conditional Laplace prior on the coefficients of interaction effect $\eta_{j}$ can be written as:
\begin{equation*}
\begin{aligned}
\pi(\eta_{j}|\tau, \lambda_2) &= \prod_{k=1}^{q}\frac{\varphi_2}{2}\text{exp}\{-\varphi_2|\eta_{jk}|\}\\
&=\prod_{k=1}^{q}\int_0^{\infty}\frac{1}{\sqrt{2\pi s_2}}\text{exp}(-\frac{\eta_{jk}^2}{2 s_2})\frac{\varphi_2^2}{2}\text{exp}(\frac{-\varphi_2^2}{2}s_2)d s_2.
\end{aligned}
\end{equation*}
Therefore, we consider the following hierarchical formulation for the marginal G$\times$E model:
\begin{equation}\label{equr:ge2}
\begin{aligned}
\beta_j|s_1, \pi_1 &{\thicksim} (1-\pi_1)\text{N}(0, s_1)+\pi_1 \delta_{0}(\beta_j),\\
s_{1}|{\varphi_{1}^{2}}&{\thicksim}\frac{\varphi_{1}^{2}}{2}\text{exp}({-\frac{\varphi_{1}^{2}}{2}}s_{1}),\\
\eta_{jk}|s_{2k}, \pi_2 &\stackrel{iid}{\thicksim} (1-\pi_2)\text{N}(0, s_{2k})+\pi_2 \delta_{0}(\eta_{jk})(k=1,\dots, q), \\
s_{2k}|{\varphi_{2}^{2}}& \stackrel{iid}{\thicksim}\frac{\varphi_{2}^{2}}{2}\text{exp}({-\frac{\varphi_{2}^{2}}{2}}s_{2k})(k=1,\dots, q),\\
\end{aligned}
\end{equation}
where $\delta_{0}(\beta_j)$ and $\delta_{0}(\eta_{jk})$ denote the spike at 0, respectively, and the slab distributions are represented by two normal distributions, $\text{N}(0, s_1)$ and $\text{N}(0, s_2k)$. Here, $\pi_1 \in [0,1]$ and $\pi_2\in [0,1]$. The mixture of the spike and slab components facilitate the selection of main and interaction effects. Instead of setting $\pi_1$ and $\pi_2$ to a fixed value such as 0.5, 
we assign conjugate beta priors on them as $\pi_1{\thicksim}\text{Beta}(r_1,u_1)$ and $\pi_2{\thicksim}\text{Beta}(r_2,u_2)$ which account for the uncertainty in $\pi_1$ and $\pi_2$. In this paper, we choose $r_1 = u_1 = r_2 = u_2 = 1$ as it gives a prior mean with 0.5 and it also allows a prior to spread out.  

In addition, the normal prior has been placed on the coefficients of environmental factor $\alpha_k (k = 1,\dots, q)$ and clinical factor $\gamma_t (t = 1, \dots, m)$ as: 
\begin{equation*}
\begin{aligned}
\alpha_k& \stackrel{iid}{\thicksim}\frac{1}{\sqrt{(2\pi \alpha_0)}}\text{exp}(-\frac{\alpha_{k}^2}{2\alpha_0})(k = 1,\dots, q) \\
\gamma_t& \stackrel{iid}{\thicksim}\frac{1}{\sqrt{(2\pi \gamma_0)}}\text{exp}(-\frac{\gamma_{t}^2}{2\gamma_0})(t = 1, \dots, m), \\
\end{aligned}
\end{equation*}
We also assume conjugate Gamma priors on $\tau$, $\varphi_1^2$ and $\varphi_2^2$ with
\begin{equation*}\label{equr:full}
\begin{aligned}
\tau &{\thicksim}\text{Gamma}(a,b),\\
\varphi_1^2&{\thicksim}\text{Gamma}(c_1,d_1),\\
\varphi_2^2&{\thicksim}\text{Gamma}(c_2,d_2).
\end{aligned}
\end{equation*}

In typical G$\times$E studies, the environmental and clinical factors are of low dimensionality and the selection of them is not of interest. Therefore, the sparsity-inducing priors have not been adopted for these factors. We consider the Bayesian LAD LASSO type of regularization in the proposed study as published studies have demonstrated that baseline penalty such as MCP and LASSO work well for marginal variable selection (\cite{SLH,CZJ}). 

It is noted that Zhang et al. (2020)  (\cite{ZSX}) has proposed a marginal sparse group MCP to respect the strong hierarchy between main and interaction effects. Their results are promising when long tailed distributions and outliers are not present in the response variable. Although sparse group (or, bi--level) variable selection has been demonstrated as being very effective in multiple G$\times$E studies based on joint models (\cite{ZFR}), in our study, there is only one group per each marginal model. The sparse group no longer has significant advantages over individual level selection. Therefore, it has not been considered here.

Our model respects the weak hierarchy of  ``main effects, interactions". If imposing the strong hierarchy is needed, the genetic factor, once it is not selected given the presence of corresponding interaction effects, can be added back to the identified marginal model for a refit to impose strong hierarchy (\cite{CZJ}). While such a practice is not uncommon in marginal interaction studies , Shi et al. (2014) (\cite{SLH}) has also revealed satisfactory performance when strong hierarchy has not been pursued.

\subsection{ The Gibbs sampler for robust marginal G$\times $E analysis }

For the $j$th genetic factor, the joint posterior distribution of all the unknown parameters conditional on data can be expressed as
\begin{equation*}\label{equr:full}
\begin{aligned}
\pi(\alpha, \gamma, \beta_{j}, \eta_{j}, v, &s_1, s_2, \tau, \varphi_1, \varphi_2, \pi_1, \pi_2, z_i| Y)\\
\propto &  \prod_{i=1}^{n} \frac{1}{\sqrt{2\pi \tau^{-1} \xi_{2}^{2} v_i}} 
\text{exp}\Big\{- \frac{(y_i - E_i \alpha - C_i \gamma - X_{ij}\beta_{j} - \tilde{W}_{i}\eta_{j})^2}{2 \tau^{-1} \xi_{2}^{2} v_i} \Big\} \\
& \times \prod_{i=1}^{n}\tau \text{exp}(- \tau v_{i})\tau^{a-1}\text{exp}(-b\tau) \frac{1}{\sqrt{2\pi}} \text{exp}(-\frac{1}{2} z_{i}^2)\\
& \times \prod_{k=1}^{q}\frac{1}{\sqrt{(2\pi \alpha_0)}}\text{exp}(-\frac{\alpha_{k}^2}{2\alpha_0})\\
& \times \prod_{t=1}^{m}\frac{1}{\sqrt{(2\pi \gamma_0)}}\text{exp}(-\frac{\gamma_{t}^2}{2\gamma_0})\\
& \times \Big( (1-\pi_1)(2\pi s_1)^{-1/2} \text{exp}(-\frac{\beta_{j}^{2}}{2 s_1})\textbf{I}_{\{\beta_{j} \neq 0\}} + \pi_1 \delta_0(\beta_{j})  \Big) \\
& \times \prod_{k=1}^{q}\Big( (1-\pi_2)(2\pi s_{2k})^{-1/2} \text{exp}(-\frac{\eta_{jk}^{2}}{2 s_{2k}})\textbf{I}_{\{\eta_{jk} \neq 0\}} + \pi_2 \delta_0(\eta_{jk})  \Big) \\
& \times \frac{\varphi_{1}^{2}}{2}\text{exp}({-\frac{\varphi_{1}^{2}}{2}}s_{1})\\
& \times \prod_{k=1}^{q} \frac{\varphi_{2}^{2}}{2}\text{exp}({-\frac{\varphi_{2}^{2}}{2}}s_{2k})\\
& \times (\varphi_{1}^{2})^{c_1-1}\text{exp}(-d_1\varphi_{1}^{2})\\
& \times (\varphi_{2}^{2})^{c_2-1}\text{exp}(-d_2\varphi_{2}^{2})\\
& \times \pi_{1}^{r_1-1}(1-\pi_1)^{u_1-1}\\
& \times \pi_{2}^{r_2-1}(1-\pi_2)^{u_2-1}\\
\end{aligned}
\end{equation*}

Let $\mu_{(-\alpha_k)}=E(y_i)-E_{ik}\alpha_k, (i=1,\dots,n), (k=1,\dots,q)$, representing the mean effect without the contribution of $E_{ik}\alpha_k$. The posterior distribution of the coefficient of environmental factor $\alpha_k$ conditional on all other parmeters can be expressed as
\begin{equation*}\label{equr:full}
\begin{aligned}
\pi(\alpha_k|& \text{rest}) \\
& \propto \pi(\alpha_k)\pi(Y|\cdot) \\
& \propto \exp\Big\{- \sum_{i=1}^{n}\frac{(y_i - E_i \alpha - C_i \gamma - X_{ij}\beta_{j} - \tilde{W}_{i}\eta_{j})^2}{2 \tau^{-1} \xi_{2}^{2} v_i} \Big\}
 \times\exp(-\frac{\alpha_{k}^2}{2 \alpha_0})\\
& \propto \exp\Big\{ -\frac{1}{2}\big[(\sum_{i=1}^{n} \frac{\tau E_{ik}^{2}}{\xi_{2}^{2} v_i}+\frac{1}{\alpha_0})\alpha_k^2 - 2\sum_{i=1}^{n}\frac{\tau(y_i-\mu_{(-\alpha_k)})E_{ik}}{\xi_2^2 v_i}\alpha_k\big]\Big\}.
\end{aligned}
\end{equation*}
Hence, the full conditional distribution of $\alpha_k$ is normal distribution $N(\mu_{\alpha_k}, \sigma_{\alpha_k}^{2})$with mean
\begin{equation*}\label{equr:meanm}
\mu_{\alpha_k} = \big(\sum_{i=1}^{n}\frac{\tau(y_i-\mu_{(-\alpha_k)})E_{ik}}{\xi_2^2 v_i}\big)\sigma_{\alpha_k}^2,
\end{equation*}
and variance
\begin{equation*}\label{equr:meanm}
\sigma_{\alpha_k}^{2} = \big( \sum_{i=1}^{n} \frac{\tau E_{ik}^2}{\xi_2^2 v_i}+\frac{1}{\alpha_0}\big)^{-1}.
\end{equation*}
The posterior distribution of the coefficient of clinical factor $\gamma_t (t=1,\dots,m)$ conditional on all other parameters can be obtained in similiar way. Let $\mu_{(-\gamma_t)}=E(y_i)-C_{it}\gamma_t,\, i=1,\dots,n$, then
\begin{equation*}\label{equr:meanm}
\gamma_t| \text{rest} {\thicksim} N(\mu_{\gamma_k}, \sigma_{\gamma_t}^{2}),
\end{equation*}
where 
\begin{equation*}\label{equr:meanm}
\begin{aligned}
\mu_{\gamma_t} &= \big(\sum_{i=1}^{n}\frac{\tau(y_i-\mu_{(-\gamma_t)})C_{it}}{\xi_2^2 v_i}\big)\sigma_{\gamma_t}^2,\\
\sigma_{\gamma_t}^{2} &= \big( \sum_{i=1}^{n} \frac{\tau C_{it}^2}{\xi_2^2 v_i}+\frac{1}{\gamma_0}\big)^{-1}.
\end{aligned}
\end{equation*}
Let $\mu_{(-\beta_j)}=E(y_i)-X_{ij}\beta_j$ and $l_1 = \pi(\beta_j=0|\text{rest})$, the conditional posterior distribution of the coefficient of genetic factor $\beta_j$ is a spike-and-slab distribution:
\begin{equation}
\beta_j|\text{rest}{\thicksim} (1-l_1) N(\mu_{\beta_j}, \sigma_{\beta_j}^2)+l_1 \delta_0(\beta_j),
\end{equation}
where 
\begin{equation*}
\begin{aligned}
\mu_{\beta_j} &= \big(\sum_{i=1}^{n}\frac{\tau(y_i-\mu_{(-\beta_j)})X_{ij}}{\xi_2^2 v_i}\big)\sigma_{\beta_j}^2,\\
\sigma_{\beta_j}^{2} &= \big( \sum_{i=1}^{n} \frac{\tau X_{ij}^2}{\xi_2^2 v_i}+\frac{1}{s_1}\big)^{-1}.
\end{aligned}
\end{equation*}
We can show that
\begin{equation*}
l_1 = \frac{\pi_1}{\pi_1 + (1-\pi_1)s_1^{-1/2}({\sigma_{\beta_j}^{2}})^{1/2}\text{exp} \{ \frac{1}{2} (\sum_{i=1}^{n}\frac{\tau(y_i-\mu_{(-\beta_j)})X_{ij}}{\xi_2^2 v_i})^2\sigma_{\beta_j}^{2}\} }.
\end{equation*}
The posterior distribution of $\beta_j$ is a mixture of a normal distribution and a point mass at 0. That is, at each iteractio of MCMC, $\beta_j$ is drawn from $N(\mu_{\beta_j}, \sigma_{\beta_j}^2)$ with probability $(1-l_1)$ and is set to 0 with probability $l_1$. \par
Similiarly, the posterior distribution of the interaction of the $j$th gene and environmental factors $\eta_{jk} (k=1,\dots,q)$ is also a spike-and-slab distribution. Denote $\mu_{(-\eta_{jk})}=E(y_i)-W_{ik}\eta_{jk}$ and $l_{2k} = \pi(\eta_{jk}=0|\text{rest})$, $\eta_{jk}$ follows this distribution:
\begin{equation}
\eta_{jk}|\text{rest}{\thicksim} (1-l_{2k}) N(\mu_{\eta_{jk}}, \sigma_{\eta_{jk}}^2)+l_{2k} \delta_0(\eta_{jk}),
\end{equation}
where 
\begin{equation*}
\begin{aligned}
\mu_{\eta_{jk}} &= \big(\sum_{i=1}^{n}\frac{\tau(y_i-\mu_{(-\eta_{jk})})\tilde{W}_{ik}}{\xi_2^2 v_i}\big)\sigma_{\eta_{jk}}^2,\\
\sigma_{\beta_j}^{2} &= \big( \sum_{i=1}^{n} \frac{\tau \tilde{W}_{ik}^2}{\xi_2^2 v_i}+\frac{1}{s_{2k}}\big)^{-1}.
\end{aligned}
\end{equation*}
And 
\begin{equation}
l_{2k} = \frac{\pi_2}{\pi_2 + (1-\pi_2)s_{2k}^{-1/2}({\sigma_{\eta_{jk}}^{2}})^{1/2}\text{exp} \{ \frac{1}{2} (\sum_{i=1}^{n}\frac{\tau(y_i-\mu_{(-\eta_{jk})})\tilde{W}_{ik}}{\xi_2^2 v_i})^2\sigma_{\eta_{jk}}^{2}\} }.
\end{equation}
The full conditional posterior distribution of $s_1$ is:
\begin{equation}\label{equr:s1}
\begin{aligned}
s_{1}&|\text{rest} \\
&\propto \pi(\beta_{j}|s_1,\pi_1) \pi(s_1|\varphi_{1}^{2})\\
&\propto\Big( (1-\pi_1)(2\pi s_1)^{-1/2} \text{exp}(-\frac{\beta_{j}^{2}}{2 s_1})\textbf{I}_{\{\beta_{j} \neq 0\}} + \pi_1 \delta_0(\beta_{j})  \Big)  \text{exp}(-\frac{\varphi_{1}^{2}}{2} s_1).
\end{aligned}
\end{equation}
When $\beta_j = 0$, equation(\ref{equr:s1}) is proportional to $\text{exp}(-\frac{\varphi_{1}^{2}}{2} s_1)$. Therefore, the posterior distribution of $s_1$ is $\text{exp}(\frac{\varphi_{1}^{2}}{2})$.\par
When $\beta_j \neq 0$, equation(\ref{equr:s1}) is proportional to 
\begin{equation*}
\begin{aligned}
\frac{1}{\sqrt{s_1}} &\text{exp}(-\frac{\varphi_{1}^{2}}{2} s_1) \text{exp}(-\frac{\beta_{j}^{2}}{2 s_1})\\
&\propto \frac{1}{\sqrt{s_1}} \text{exp} \Big\{ -\frac{1}{2}[\varphi_{1}^{2} s_1 + \frac{\beta_{j}^{2}}{s_1} ]  \}.
\end{aligned}
\end{equation*}
Therefore, when $\beta_j \neq 0$, the posterior distribution for $s_1^{-1}$ is Inverse-Gaussian$(\sqrt{\frac{\varphi_1^2}{\beta_j^2}}, \varphi_1^2)$.\par
Similiarly, for $s_{2k} (k=1,\dots,q)$, when $\eta_{jk} = 0$, the posterior distribution of $s_{2k}$ is $\text{exp}(\frac{\varphi_{2}^{2}}{2})$. When $\eta_{jk} \neq 0$, the posterior distribution for $s_{2k}^{-1}$ is Inverse-Gaussian$(\sqrt{\frac{\varphi_2^2}{\eta_{jk}^2}}, \varphi_2^2)$.\par
The full conditional posterior distribution of $\varphi_1^2$:
\begin{equation*}
\begin{aligned}
\varphi_1^2&|\text{rest} \\
&\propto \pi(s_1|\varphi_1^2)\pi(\varphi_1^2)\\
&\propto \frac{\varphi_1^2}{2} \text{exp}(-\frac{\varphi_1^2 s_1}{2})(\varphi_1^2)^{c_1-1}\text{exp}(-d_1\varphi_1^2)\\
&\propto (\varphi_1^2)^{c_1} \text{exp}\Big(- \varphi_1^2(s_1/2+d_1)\Big).
\end{aligned}
\end{equation*}
Therefore, the posterior distribution for $\varphi_1^2$ is Gamma($c_1+1, s_1/2+d_1$).
Similiarly, the posterior distribution for $\varphi_2^2$ is Gamma($c_2+q, \sum_{k=1}^{q} s_{2k}/2+d_2$).\par
The full conditional posterior distribution of $\pi_1$:
\begin{equation*}
\begin{aligned}
\pi_1&|\text{rest} \\
&\propto \pi(s_1|\varphi_1^2)\pi(\varphi_1^2)\\
&\propto \pi_{1}^{r_1-1}(1-\pi_1)^{u_1-1}\\
& \times \Big( (1-\pi_1)(2\pi s_1)^{-1/2} \text{exp}(-\frac{\beta_{j}^{2}}{2 s_1})\textbf{I}_{\{\beta_{j} \neq 0\}} + \pi_1 \delta_0(\beta_{j})  \Big).
\end{aligned}
\end{equation*}
Then, the posterior distribution for $\pi_1$ is Beta ($1+r_1- \textbf{I}{(\beta_j \neq 0)}, u_1+\textbf{I}{(\beta_j \neq 0)}$).\par
The full conditional posterior distribution of $\pi_2$:
\begin{equation*}
\begin{aligned}
\pi_2&|\text{rest} \\
&\propto \pi(s_2|\varphi_2^2)\pi(\varphi_2^2)\\
&\propto\pi_{2}^{r_2-1}(1-\pi_2)^{u_2-1}\\
& \times \prod_{k=1}^{q}\Big( (1-\pi_2)(2\pi s_{2k})^{-1/2} \text{exp}(-\frac{\eta_{jk}^{2}}{2 s_{2k}})\textbf{I}_{\{\eta_{jk} \neq 0\}} + \pi_2 \delta_0(\eta_{jk})  \Big).
\end{aligned}
\end{equation*}
So, the posterior distribution for $\pi_2$ is Beta $(1+r_1-\sum_{k=1}^{q}\textbf{I}{(\eta_{jk} \neq 0)}, u_1+\sum_{k=1}^{q}\textbf{I}{(\eta_{jk} \neq 0)})$.\par
The full conditional posterior distribution of $\tau$:
\begin{equation*}
\begin{aligned}
\tau&|\text{rest} \\
& \propto \pi(v|\tau)\pi(\tau)\pi(Y|\cdot) \\
& \propto {\tau}^{n/2}\exp\Big\{- \sum_{i=1}^{n}\frac{(y_i - E_i \alpha - C_i \gamma - X_{ij}\beta_{j} - \tilde{W}_{i}\eta_{j})^2}{2 \tau^{-1} \xi_{2}^{2} v_i} \Big\}\\
& \times {\tau}^n \text{exp}(-\tau\sum_{i=1}^{n} v_i) \tau^{a-1} \text{exp}(-b\tau)\\
&\propto \tau^{a+\frac{3}{2}n-1} \text{exp}\Big\{-\tau\big[\sum_{i=1}^{n}(\frac{(y_i - E_i \alpha - C_i \gamma - X_{ij}\beta_{j} - \tilde{W}_{i}\eta_{j})^2}{2  \xi_{2}^{2} v_i} +v_i)+b\big] \Big\}.
\end{aligned}
\end{equation*}
Therefore, the posterior distribution for $\tau$ is Gamma($a+\frac{3}{2}n, \:\big[\sum_{i=1}^{n}(\frac{(y_i - E_i \alpha - C_i \gamma - X_{ij}\beta_{j} - \tilde{W}_{i}\eta_{j})^2}{2  \xi_{2}^{2} v_i} +v_i)+b\big]$).\par
Last, we have the full conditional posterior distribution of $v_i$:
\begin{equation*}
\begin{aligned}
v_i&|\text{rest} \\
& \propto \pi(v|\tau)\pi(Y|\cdot) \\
& \propto \frac{1}{\sqrt{v_i}} \exp\Big\{- \frac{(y_i - E_i \alpha - C_i \gamma - X_{ij}\beta_{j} - \tilde{W}_{i}\eta_{j})^2}{2 \tau^{-1} \xi_{2}^{2} v_i} \Big\}
\times\exp(-\tau v_i)\\
& \propto \frac{1}{\sqrt{v_i}} \exp\Big\{-\frac{1}{2} \big[ (2\tau)v_i + \frac{\tau(y_i - E_i \alpha - C_i \gamma - X_{ij}\beta_{j} - \tilde{W}_{i}\eta_{j})^2}{\xi_{2}^{2} v_i}\big] \Big\}.
\end{aligned}
\end{equation*}
It is easy to show that 
\begin{equation*}
\frac{1}{v_i}|\text{rest} {\thicksim}\text{ Inverse-Gaussian} (\sqrt{\frac{2 \xi_2^2}{(y_i - E_i \alpha - C_i \gamma - X_{ij}\beta_{j} - \tilde{W}_{i}\eta_{j})^2}}, \:2\tau).
\end{equation*}

The spirit of marginal penalization for G$\times$E interactions lies in the usage of a common sparsity cutoff to determine a list of important main and interaction effects. Instead of focusing on a fixed cutoff, varying the cutoff can generate different lists, resulting in a comprehensive view of important findings. The tuning parameter in penalized estimation serves as the cutoff. Therefore, the same tuning parameter has to be adopted for all the sub models (\cite{SLH,CZJ,ZSX}). To further justify such a common tuning parameter, Zhang et al. (2020) (\cite{ZSX}) has attempted using the joint model to select the common tuning through cross validation. However, this seems not coherent with the nature of marginal analysis. 

Ideally, the tuning parameter should be determined by each model itself to allow for flexibility in controlling sparsity individually, and a common cutoff is still available to examine different lists of important effects. With the Bayesian formulation, we can avoid such a limitation of frequentist marginal penalization methods. In particular, the priors have been placed on regularization parameters to determine the sparsity in a data-driven manner for each sub model. With the spike-and-slab priors, the posterior distributions on the coefficients of main and interaction effects naturally lead to the usage of inclusion probability as a common cutoff to pin down the list of important effects, which is described in detail in the next section.

\section{Simulation}

To demonstrate the utility of the proposed approach, we evaluate the performance through simulation study. In particular, we compare the performance of the proposed method, LAD Bayesian Lasso with spike-and-slab priors (denoted as LADBLSS) with three alternatives, LAD Bayesian Lasso (denoted as LADBL), Bayesian Lasso with spike-and-slab priors (denoted as BLSS) and Bayesian Lasso (denoted as BL). LADBL is similar to the proposed method, except that it does not adopt the spike-and-slab prior. The details of posterior inference are available from the Appendix.

Under all settings, the sample size is set as $n=200$, and the number of G factors is $p=500$ with $q=4$, $m=3$. For environmental factors, we simulate four continuous variables from multivariate normal distributions with marginal mean 0, marginal variance 1 and AR1 correlation structure with $\rho=0.5$. In addition, three clinical factors are generated from a multivariate normal distribution with margianl mean 0 and marginal variance 1 and AR1 structure with $\rho=0.5$.  Among the $p$ main G effects and $pq$ G$\times $E interactions, 8 and 12 effects are set as being associated with the response, respectively. All the environmental and clinical factors are important with nonzero coefficients, which are randomly generated from a uniform distribution Unif$[0.1, 0.5]$. The random error are generated from: (1) N$(0,1)$(Error 1), (2) t-distribution with 2 degress of freedom ($t$(2)) (Error2), (3) LogNormal(0,2)(Error3), (4) 90\%N(0,1)+10\%Cauchy(0,1)(Error4), (5) 80\%N(0,1)+20\%Cauchy(0,1)(Error5). All of them are heavy-tailed distribution except the first one. 

In addition, the genetic factors are simulated in the following four settings.

\textit{Setting 1}. In simulating continuous genetic variants, we generate multivariate normal distributions with marginal mean 0 and variance 1. The AR structure is considered in computing the correlation of G factors, under which gene $j$ and $k$ have correlation $\rho^{|j-k|}$ with $\rho=0.5$.

\textit{Setting 2}. We assess the performance under single-nucleotide polymorphism (SNP) data. The SNPs are obtained by dichotomizing the gene expression values at the 1st and 3rd quartiles, with the 3--level (0,1,2) for genotypes (aa,Aa,AA) respectively. Here, the gene expressions are generated from the first setting.

\textit{Setting 3}. Consider simulating the SNP data under a pairwise linkage disequilibrium (LD) structure. For the two minor alleles A and B of two adjacent SNPs, let $q_{1}$ and $q_{2}$ be the minor allele frequencies (MAFs), respectively. The frequencies of four haplotypes are as $p_{AB} = q_{1}q_{2} + \delta$, $p_{ab} = (1-q_{1})(1-q_{2})+\delta$, $p_{Ab} = q_{1}(1-q_{2})-\delta$, and $p_{aB} = (1-q_{1})q_{2}-\delta$, where $\delta$ denotes the LD. Assuming Hardy-Weinberg equilibrium and given the allele frequency for A at locus 1, we can generate the SNP genotype (AA, Aa, aa) from a multinomial distribution with frequencies $(q^{2}_{1}, 2q_{1}(1-q_{1}),(1-q_{1})^{2})$. Based on the conditional genotype probability matrix, we can simulate the genotypes for locus 2. With MAFs 0.3 and pairwise correlation $r=0.6$, we have $\delta=r\sqrt{q_{1}(1-q_{1})q_{2}(1-q_{2})}$.

We collect the posterior samples from the Gibbs Sampler with 10,000 interations and discard the first 5,000 samples as burn-ins. The posterior medians are used to estimate the coefficients. For approaches incorporating spike-and-slab priors, we consider computing the inclusion probability to indicate the importance of predictors. Here we use a binary indicator $\phi$ to denote that the membership of the non-spike distribution. Take the main effect of the $j$th genetic factor, $X_{j}$, as an example. Suppose we have collected H posterior samples from MCMC after burn-ins. The $j$th G factor is included in the marginal G$\times$E model at the $h$th MCMC iteration if the corresponding indicator is 1, i.e., $\phi_j^{(h)} = 1$. Subsequently, the posterior probability of retaining the $j$th genetic main effect in the final marginal model is defined as the average of all the indicators for the $j$th G factor among the H posterior samples. That is,
\begin{equation*}
p_j = \hat{\pi} (\phi_j = 1|y) = \frac{1}{H} \sum_{h=1}^{H} \phi_j^{(h)}, \;  j = 1, \dots,p.
\end{equation*}
A larger posterior inclusion probability $p_j$ indicates a stronger empirical evidence that the $j$th genetic main effect has a non-zero coefficient, i.e., a stronger association with the phenotypic trait. 

To comprehensively assess the performance of the proposed and alternative methods, we consider a sequence of probabilities as cutting-offs in inclusion probability for methods with spike-and-slab priors. Given a cutoff probability, the main or interaction is included in the final marginal model if its posterior inclusion probability is larger than the cutoff, and is excluded otherwise. Provided with a sequence of cutting-off probabilities from small to large, we can investigate the set of identified effects and calculate the true/false positive rates (T/FPR) as the ground truth is known in simulation. For the sequence of cut-offs, we are able to compute the area under curve (AUC) as a comprehensive measure. Besides, for methods without spike-and-slab priors, the confidence level of the credible intervals can be adopted as the cut-off to compute TPR and FPRs. Therefore, all the methods under comparison can be evaluated on the same ground.

In addition, we also consider Top100, which is defined as the number of true signals when 100 important main effects (or interactions) are identified. For methods with spik-and-slab priors, 100 main effects or interactions are chosen with the highest inclusion probabilities. For methods without spike-and-slab priors, the indicators of all effects are computed for a sequence of credible levels. The top 100 main effects or interactions are chosen in terms of the highest average identification values.


Simulation results for the gene expression data in the first setting are tabulated in Tables \ref{id1.1}and \ref{id1.2}. We can observe that the proposed method has the best performance among all approaches, especially when the response variable has heavy-tailed distributions. First, the performance of methods with spike-and-slab priors is consistently better than methods without spike-and-slab priors. For example, in Table \ref{id1.1}, under error 3, the AUC of LADBLSS is 0.9558(sd 0.0161), which is much larger than that of the robust method without spike-and-slab priors, i.e., 0.8432(sd 0.0115) from LADBL. Also, the AUC of robust methods is much larger than that of non-robust methods, especially in the presence of heavy-tailed errors. For instance, in the first setting under error3, the AUC of LADBLSS is 0.9558 and the AUC of LADBL is 0.8432 while that of BLSS and BL is around 0.5. Similar advantageous performance can also be observed from the identification results with Top100.  In Table \ref{id1.2} under error 5, LADBLSS identifies 7.80(sd 0.55) out of the 8 main effects and 10.53(sd 1.36) out of the 12 interaction effects. This is higher than the results of LADBL with 7.57(sd 0.57) of main effects and 6.83(sd 1.07) of interaction effects. Second, among all the methods with spike-and-slab priors,  Bayesian LAD method with spike-and-slab priors has the best performance in all identification results. Under error 3, in Table \ref{id1.1}, the AUC of LADBLSS is 0.9558(sd 0.0161) while the AUC of BLSS is 0.5473(sd 0.0576). Under error 4 in Table \ref{id1.2}, LADBLSS identifies 7.77(sd 0.57) main effects and 10.67(sd 1.50) interaction effects while BLSS identifies 6.2(sd 2.62) main effects and 8.3(sd 3.98) interaction effects, respectively.

\begin{table} [h!]
\def\arraystretch{1.7}
\begin{center}
\caption{Simulation results of the first setting. AUC (mean of AUC), SD (sd of AUC) based on 100 replicates. $n$=200, $p$=500, $q$=4 and $m=3$. }\label{id1.1}
\centering
\fontsize{10}{10}\selectfont{
\begin{tabular}{ c c c c c c c c  }
\hline
   &     &  BL  & BLSS  & LADBL   & LADBLSS    \\
\hline
Error 1  & AUC    &0.9182 &0.9901  &0.9258	&0.9887  \\
  N(0,1)         & SD   & 0.0052 &0.0021 &0.0076 &0.0026\\
\hline
Error 2              & AUC    &0.8332 & 0.9420 & 0.9004 & 0.9841  \\
 $t$(2)                    & SD   &0.0107 & 0.0235 & 0.0078 & 0.0031  \\

\hline
Error 3    & AUC    & 0.5343 & 0.5473 & 0.8432 & 0.9558 \\
 Lognormal(0,2)          & SD   &0.0144 & 0.0576 & 0.0115  & 0.0161 \\
\hline
Error 4           & AUC   & 0.8221 &  0.9124 & 0.9222 & 0.9895 \\
90$\%$N(0,1)+10$\%$Cauchy(0,1)                  & SD    & 0.0212 & 0.0410 &  0.0071 & 0.0024 \\
\hline
Error 5    &AUC    & 0.7507 & 0.8431	 & 0.9192 & 0.9904 \\
 80$\%$N(0,1)+20$\%$Cauchy(0,1)           & SD    & 0.0217 & 0.0633 & 0.0059 & 0.0018 \\
\hline 
\end{tabular} }
\end{center}
\centering
\end{table}

\begin{table} [ht!]
\def\arraystretch{1.5}
\begin{center}
\caption{Identification results of the first setting with Top100 method. mean(sd) based on 100 replicates. $n$=200, $p$=500, $q$=4 and $m=3$. }\label{id1.2}
\centering
\fontsize{10}{10}\selectfont{
\begin{tabular}{ c c c c c c c c  }
\hline
    &     & Main   & Interaction  & Total   \\
\hline
Error 1  & BL    &7.60(0.49) &6.80(1.6)  &14.40(1.73)  \\
   N(0,1)          & BLSS    &7.80(0.41) & 10.80(0.92) & 18.60(1.13)\\ 
     & LADBL    & 7.67(0.55) &6.53(1.85)  &14.20(1.81) \\
         & LADBLSS    &7.76(0.5) &10.53(1.36)  &18.30(1.49) \\
\hline
Error 2              & BL    &6.37(1.90) &3.90(2.07)  &10.27(3.19) \\
 $t$(2)            & BLSS    &6.33(1.63) &8.53(2.46)  &14.87(3.71) \\      
                & LADBL    &7.43(0.94) &5.80(1.71)  &13.23(2.01) \\
                      & LADBLSS    &7.53(0.51) &9.90(1.56)  &17.43(1.76) \\
\hline
Error 3    & BL    &0.90(1.21) &0.50(0.97)  &1.40(1.45) \\
 Lognormal(0,2)   & BLSS &0.73(0.94) &0.47(0.68)  &1.20(1.35) \\  
          & LADBL   &6.27(1.55) &3.67(1.94)  &9.93(2.75) \\
           & LADBLSS    &6.10(1.37) &8.93(2.02) &15.03(3.09) \\
\hline
Error 4           & BL    &5.57(2.99) &3.63(2.53)  &9.20(5.05) \\
 90$\%$N(0,1)            & BLSS    &6.20(2.62) &8.30(3.98)  &14.50(6.39) \\
+10$\%$Cauchy(0,1)   & LADBL    &7.77(0.43) &7.00(1.93)  &14.77(1.81) \\
                    & LADBLSS   & 7.77(0.57) &10.67(1.50)  &18.23(1.67) \\
\hline
Error 5           & BL    & 5.07(2.89)&3.00(2.49)  &8.07(5.01) \\
 80$\%$N(0,1)                 & BLSS    & 4.60(3.25) &5.70(4.23)  &10.30(7.27) \\
+20$\%$Cauchy(0,1)   & LADBL    & 7.57(0.57) & 6.83(1.07) &14.40(1.83) \\
                    & LADBLSS    & 7.80(0.55) &10.53(1.36)  &18.33(1.69) \\
\hline 
\end{tabular} }
\end{center}
\centering
\end{table}

\begin{figure}[h!]
	\centering

	\includegraphics[angle=0,origin=c,width=0.9\textwidth]{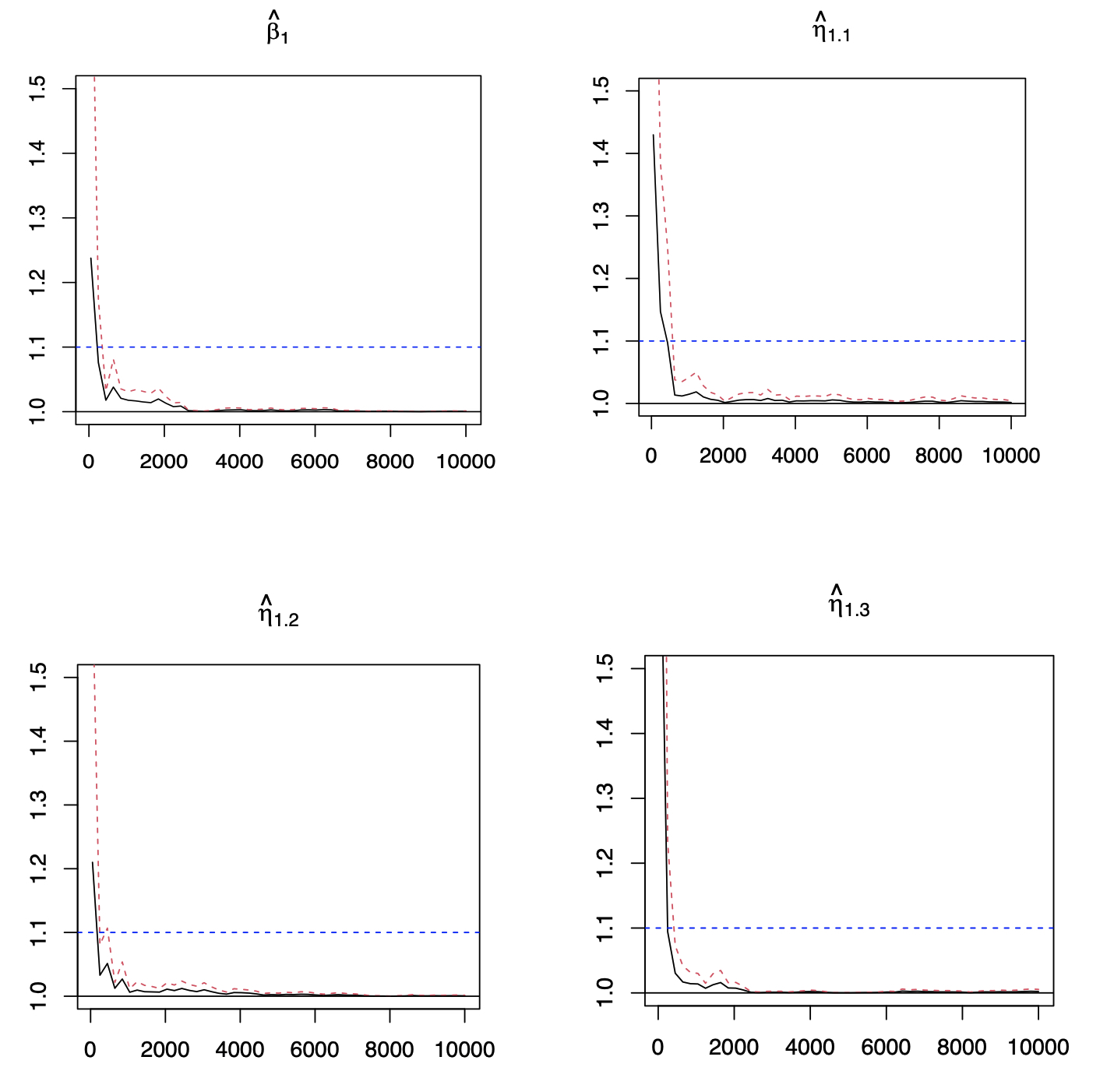}
	\caption[Potential scale reduction factor against iterations]{{Potential scale reduction factor (PSRF) against iterations for the coefficients of the first genetic factors and its interaction with environmental factors in Example 1 under Error 3. Black line: the PSRF. 
			Red dotted line: the upper limits of the 95\%  confidence interval for the PSRF. 
Blue dotted line: the threshold of 1.1. The $\hat{\beta}_{1}$ represents the estimated coefficients of the main effects for the first genetic factor. The $\hat{\eta}_{11}$ to $\hat{\eta}_{13}$ represent the estimated coefficients of the first three interaction effects for the first genetic factor.}}
	\label{fig:psrf}
\end{figure}

Similar patterns can be observed in Table \ref{id2.1}, \ref{id2.2} for the second setting, and Table \ref{id3.1}, \ref{id3.2} for the third setting in Appendix. Overall, the advantages of conducting robust Bayesian G$\times$E analysis using the proposed approach can be justified based on the results of comprehensive simulation studies. The convergence of the MCMC chains with the potential scale reduction factor (PSRF) (\cite{BROOK}) has been conducted. In this study, we use PSRF $\leq$ 1.1 (\cite{GEL}) as the cut-off point which indicates that chains converge to a stationary distribution. The convergence of chains after burn-ins has been checked for all parameters with the value of PSRF less than 1.1. Figure \ref{fig:psrf} shows the convergence pattern of PSRF for the main and interaction coefficients of the first genetic factors in Example 1 under Error 3.

\section{Real Data Analysis}
In this study, we analyze the type 2 diabetes (T2D) data from Nurses' Health Study (NHS), which is a well-characterized cohort study of women with high dimensional SNP data, as well as measurements on lifestyle and dietary factors. We consider SNPs on chronmosome 10 to identify main and gene-environment interactions associated with weight, which is an important phenotypic trait related to type 2 diabetes. Here, weight is used as response and five environment factors, age (age), total physical activity (act), trans fat intake (trans), cereal fiber intake (ceraf) and reported high blood cholesterol (chol) are considered. Data are available on 3391 subjects and 17016 gene expressions after cleaning the raw data through matching phenotypes and genotypes and removing SNPs with minor allele frequency (MAF) less than 0.05. A prescreening is done before downstream analysis. We use a marginal linear model with weight as response and age, act, trans, ceraf, chol as environment factors. 10,000 SNPs which have at least two main or interaction effects with p-value less than 0.05 are kept.

We use Top 100 method to identify 100 most important main and interaction effects. The proposed method LADBLSS identifies 20 main SNP effects and 80 gene-environment interactions, which are listed in Table \ref{T2D.LADBLSS}. Our study provides crucial implications in identifying the important main and interactions of SNPs and its associations with weight. For example, three SNPs, rs17011106, rs4838643 and rs17011115, located within gene WDFY4 are identified. WDFY4 has been observed as an influential factor related to weight and obesity (\cite{BAR, MARTIN}). In addition, SNPs rs10994364, rs10821773 and rs10994308, located within gene ANK3, are identified with interacting environment factors age and chol. There are findings showing an association between ANK3 and higher systolic blood pressure (\cite{GHAN}). Published studies have also shown that ANK3 is linked to plumonary and renal hypertension (\cite{GHAN}). Allele risk variants have been identified in ANK3, and these variants explain a proportion of the heritability of BD (bipolar disorder), which is associated with higher body mass index (BMI) and increased metabolic comorbidity and the genetic risk for BD relates to common genetic risk with T2D (\cite{WIN}). Our proposed method identifies its interaction with chol, the high blood cholesterol.Data from several sources suggest that islet cholesterol metabolism contributes to the pathogenesis of T2D (\cite{BRUN}). 

Analysis with alternatives BL, BLSS and LADBL has also been conducted. To compare the alternative methods with the proposed method, we provide the numbers of main effects and interactions identified by these methods with pairwise overlaps in Table \ref{case1.1}. It clearly shows that the proposed one results in a very different set of effects compared to alternatives. We refit the regularized marginal models by LADBL and LADBLSS using robust Bayesian Lasso, and those identified by BL and BLSS using Bayesian Lasso. In addition, the inclusion probabilities of the selected main and interaction effects using LADBLSS are provided in Table \ref{INCLU.LADBLSS}. Results from the alternative methods are available from the Supplementary files. The proposed method selects the 100 most important effects with the inclusion probability larger than 0.9, which shows advantage in quantifying uncertain compared to marginal penalization methods (\cite{SLH,CZJ,ZSX}).


\begin{table} [ht!]
\def\arraystretch{1.7}
\begin{center}
\caption{The numbers of main G effects and interactions identified by different approaches and their overlaps.}\label{case1.1}
\centering
\fontsize{10}{10}\selectfont{
\begin{tabular}{ c c c c c| c c c c  }

\hline
 \textbf{T2D}      & \multicolumn{4}{c}{Main}  & \multicolumn{4}{c}{Interaction} \\
\cline{2-5} \cline{6-9}
		
         & BL & BLSS   & LADBL  & LADBLSS  & BL & BLSS   & LADBL  & LADBLSS   \\
\hline
          BL& 86 &5 &6  &8 &14 &14 &4 &8 \\
          BLSS&  &24 & 3 &6 & &76&20&23 \\
          LADBL& & &20 &12 & & &80&50 \\
          LADBLSS& & &  &20 & & & &80 \\   
\hline 

\end{tabular} }
\end{center}
\centering
\end{table}

\newpage
\section{Discussion}

In the past, G$\times$E interaction studies have been mainly conducted through marginal hypothesis testing, based on a diversity of study designs utilizing parametric, nonparametric and semiparametric models (\cite{MURC, THOM, MUKB}), which later have been extended to joint analyses driven primarily by the pathway or gene set based association studies (\cite{WUCU, JZS, JHDZ}). In addition, published literature has also reported the success of marginal screening studies, including those based on partial correlations (\cite{XWZ, NHZ}).
Recently, the effectiveness of regularized variable selection in G$\times$E interaction studies has been increasingly recognized, and a large number of regularization methods have been proposed for joint interaction studies (\cite{ZFR}). Marginal penalization has also been demonstrated as promising competitors, although they have only been investigated in a limited number of frequentist studies (\cite{SLH, ZSX, CZJ}).

Therefore, the proposed marginal robust Bayesian variable selection is of particular importance, since joint and marginal analysis cannot replace each other and marginal Bayesian penalization has not been examined for G$\times$E studies so far. In particular, with the robustness and incorporation of spike- and-slab priors in the adaptive Bayesian shrinkage, the LADBLSS has an analysis framework more coherent with that of the joint robust analysis (\cite{RZLM}), which significantly facilitates methodological developments for interaction studies. 

The marginal Bayesian regularization can be extended to different types of response, for example, under binary, categorical, prognostic and multivariate outcomes. Nevertheless, considering robustness in the generalized models with the Bayesian framework is not trivial, especially under the multivariate responses (\cite{WCM, ZRLJ}). We postpone the investigations to the future studies.The interaction between genetic and environmental factors in this study has been modeled as the product of the two corresponding variables, which amounts to “linear” interactions. In practice, the linear interaction assumption has been frequently violated(\cite{MAYR, WUCY,ZZC}), which demands accommodation of these nonlinear effects through nonparametic and semiparametric models (\cite{LWL, RZL, WZCY, WSCM}). It is of great interest and importance to migrate the nonlinear G$\times$E studies to marginal cases in the near future.


\appendix
\section{Additional simulation results}
\subsection{Identification results in simulation}

\begin{table} [h!]
\def\arraystretch{1.7}
\begin{center}
\caption{Simulation results of the second setting. AUC (mean of AUC), SD (sd of AUC) based on 100 replicates. $n$=200, $p$=500, $q$=4 and $m=3$. }\label{id2.1}
\centering
\fontsize{10}{10}\selectfont{
\begin{tabular}{ c c c c c c c c  }
\hline
   &     &  BL  & BLSS  & LADBL   & LADBLSS    \\
\hline
Error 1  & AUC    &0.9089 &0.9881  &0.9148	&0.9888  \\
N(0,1)         & SD   & 0.0059 &0.0019 &0.0051 &0.0037\\
\hline
Error 2              & AUC    &0.8187 & 0.9255 & 0.8877& 0.9769  \\
$t$(2)                     & SD   &0.0142 & 0.0524 & 0.0057 & 0.0048  \\

\hline
Error 3    & AUC    & 0.5333 & 0.5533 & 0.8239 & 0.9459 \\
Lognormal(0,2)           & SD   &0.0096 & 0.0656 & 0.1045  & 0.0162 \\
\hline
Error 4           & AUC   & 0.8113 &  0.9122 & 0.9111 & 0.9849 \\
90$\%$N(0,1)+10$\%$Cauchy(0,1)                  & SD    & 0.0166 & 0.0502 &  0.0083 & 0.0033 \\
\hline
Error 5    &AUC    & 0.7425 & 0.8086 & 0.9076 & 0.9856 \\
80$\%$N(0,1)+20$\%$Cauchy(0,1)           & SD    & 0.0241 & 0.0746 & 0.0065 & 0.0024 \\
\hline 
\end{tabular} }
\end{center}
\centering
\end{table}

\begin{table} [ht!]
\def\arraystretch{1.5}
\begin{center}
\caption{Identification results of the second setting with Top100 method. mean(sd) based on 100 replicates. $n$=200, $p$=500, $q$=4 and $m=3$. }\label{id2.2}
\centering
\fontsize{10}{10}\selectfont{
\begin{tabular}{ c c c c c c c c  }
\hline
    &     & Main   & Interaction  & Total   \\
\hline
Error 1  & BL    &7.50(0.86) &6.70(1.49)  &14.20(1.83)  \\
   N(0,1)          & BLSS    &7.60(0.67) & 10.20(0.09) & 17.80(1.32)\\ 
     & LADBL    & 7.67(0.66) &6.83(1.82)  &14.5(1.96) \\
         & LADBLSS    &7.63(0.56) &9.97(1.54)  &17.6(1.67) \\
\hline
Error 2              & BL    &5.83(2.21) &3.47(1.57)  &9.30(2.98) \\
 $t$(2)            & BLSS    &6.33(2.09) &7.57(3.15)  &13.90(4.73) \\      
                & LADBL    &7.07(0.94) &5.97(1.61)  &13.03(1.96) \\
                      & LADBLSS    &7.40(0.62) &9.20(1.94)  &16.60(2.11) \\
\hline
Error 3    & BL    &0.77(0.86) &0.73(0.94)  &1.50(1.11) \\
 Lognormal(0,2)   & BLSS &0.57(1.01) &0.67(1.06)  &1.23(1.77) \\  
          & LADBL   &5.90(1.65) &3.50(1.96)  &9.40(2.43) \\
           & LADBLSS    &5.67(1.73) &9.00(2.35) &14.67(3.73) \\
\hline
Error 4           & BL    &6.03(2.19) &4.40(2.44)  &10.43(4.17) \\
 90$\%$N(0,1)            & BLSS    &6.03(2.57) &8.00(3.33)  &14.03(5.76) \\
+10$\%$Cauchy(0,1)   & LADBL    &7.27(0.91) &6.87(1.48)  &14.13(1.74) \\
                    & LADBLSS   & 7.53(0.63) &10.00(1.43)  &17.53(1.57) \\
\hline
Error 5           & BL    & 5.53(2.45)&3.63(2.19)  &9.16(4.13) \\
 80$\%$N(0,1)                 & BLSS    & 5.07(2.57) &6.73(3.37)  &11.80(5.65) \\
+20$\%$Cauchy(0,1)   & LADBL    & 7.47(0.97) & 5.43(1.77) &12.90(2.04) \\
                    & LADBLSS    & 7.37(0.85) &10.47(1.46)  &17.83(1.91) \\
\hline 
\end{tabular} }
\end{center}
\centering
\end{table}

\begin{table} [h!]
\def\arraystretch{1.7}
\begin{center}
\caption{Simulation results of the third setting. AUC (mean of AUC), SD (sd of AUC) based on 100 replicates. $n$=200, $p$=500, $q$=4 and $m=3$. }\label{id3.1}
\centering
\fontsize{10}{10}\selectfont{
\begin{tabular}{ c c c c c c c c  }
\hline
   &     &  BL  & BLSS  & LADBL   & LADBLSS    \\
\hline
Error 1  & AUC    &0.9158 &0.9895  &0.9251	&0.9878  \\
N(0,1)         & SD   & 0.0041 &0.0022 &0.0054 &0.0028\\
\hline
Error 2              & AUC    &0.8323 & 0.9461 & 0.8972& 0.9833  \\
 $t$(2)                     & SD   &0.0117 & 0.0342 & 0.0062 & 0.0028  \\

\hline
Error 3    & AUC    & 0.5268 & 0.5531 & 0.8415 & 0.9595 \\
Lognormal(0,2)           & SD   &0.0127 & 0.0590 & 0.0107  & 0.0156 \\
\hline
Error 4           & AUC   & 0.8261 &  0.9323 & 0.9245 & 0.9889 \\
 90$\%$N(0,1)+10$\%$Cauchy(0,1)                 & SD    & 0.0191 & 0.0352 &  0.0056 & 0.0034 \\
\hline
Error 5    &AUC    & 0.7533 & 0.8591 & 0.9204 & 0.9862 \\
80$\%$N(0,1)+20$\%$Cauchy(0,1)            & SD    & 0.0201 & 0.0657 & 0.0067 & 0.0114 \\
\hline 
\end{tabular} }
\end{center}
\centering
\end{table}

\begin{table} [ht!]
\def\arraystretch{1.5}
\begin{center}
\caption{Identification results of the third setting with Top100 method. mean(sd) based on 100 replicates. $n$=200, $p$=500, $q$=4 and $m=3$. }\label{id3.2}
\centering
\fontsize{10}{10}\selectfont{
\begin{tabular}{ c c c c c c c c  }
\hline
    &     & Main   & Interaction  & Total   \\
\hline
Error 1  & BL    &7.70(0.47) &6.80(1.63)  &14.50(1.79)  \\
   N(0,1)          & BLSS    &7.63(0.72) & 10.93(0.98) & 18.57(1.22)\\ 
     & LADBL    & 7.70(0.75) &7.33(1.95)  &15.03(2.14) \\
         & LADBLSS    &7.87(0.35) &10.33(1.35)  &18.20(1.45) \\
\hline
Error 2              & BL    &6.57(1.87) &4.47(1.69)  &11.03(2.88) \\
 $t$(2)            & BLSS    &6.60(1.57) &8.40(2.51)  &15.00(3.68) \\      
                & LADBL    &7.57(0.62) &5.77(1.50)  &13.33(1.77) \\
                      & LADBLSS    &7.43(0.68) &9.30(2.15)  &16.73(2.43) \\
\hline
Error 3    & BL    &0.50(0.73) &0.83(1.02)  &1.33(1.47) \\
 Lognormal(0,2)   & BLSS &0.70(0.99) &0.40(0.86)  &1.10(1.54) \\  
          & LADBL   &6.13(2.05) &3.80(1.39)  &9.93(1.32) \\
           & LADBLSS    &6.63(1.16) &10.10(1.73) &16.73(2.52) \\
\hline
Error 4           & BL    &5.73(2.82) &4.30(2.64)  &10.03(5.11) \\
 90$\%$N(0,1)            & BLSS    &5.73(3.02) &7.67(4.19)  &13.40(7.05) \\
+10$\%$Cauchy(0,1)   & LADBL    &7.80(0.48) &6.87(1.61)  &14.67(1.54) \\
                    & LADBLSS   & 7.83(0.38) &10.50(1.25)  &18.33(1.39) \\
\hline
Error 5           & BL    & 5.60(2.61)&2.93(2.23)  &8.53(4.27) \\
 80$\%$N(0,1)                 & BLSS    & 5.27(2.27) &6.90(3.64)  &12.17(5.66) \\
+20$\%$Cauchy(0,1)   & LADBL    & 7.87(0.35) & 6.87(1.45) &14.73(1.46) \\
                    & LADBLSS    & 7.70(0.53) &10.70(1.12)  &18.40(1.28) \\
\hline 
\end{tabular} }
\end{center}
\centering
\end{table}

\clearpage
\section{{Assessment of the convergence of MCMC chains}}

\clearpage
\section{Estimation results for data analysis}

\begin{longtable}{ c c c c c c c c c}

\caption[Analysis of the NHS T2D data using LADBLSS.]{Analysis of the NHS T2D data using LADBLSS.} \label{T2D.LADBLSS} \\
\hline
   &    &     &    &   & Interactions   &  &  &   \\
\cline{4-9}
SNP  &Gene      & Main Effects   & age   & act  & trans   & ceraf & chol \\
\hline	
	\endfirsthead
	\caption[]{Continued from the previous page.} \\ \hline
   &    &     &    &   & Interactions   &  &  &   \\
\cline{4-9}
SNP  &Gene      & Main Effects   & age   & act  & trans   & ceraf & chol \\
\hline	

	\endhead
	\hline 
	\multicolumn{8}{r@{}}{Continued on the next page}\\
	\endfoot
	\hline
	\endlastfoot
rs17011106      & WDFY4 &-0.024  & & &&&\\
rs7077294 &KIAA1217 & &&&&&-0.0491\\
rs7093682 &RP11-170M17.1& & & &-0.1239 &&\\
rs17011106 &WDFY4&-0.0953 & & & &&\\
rs10826028 &MIR3924&  & & & &-0.0524&\\
rs4748996 &THNSL1&  & & & & &0.0064\\
rs2646392 &KRT8P37&0.0148  & & & & & \\
rs7904629 &RP11-170M17.1& & & &-0.0592 & & \\
rs1244416 &ATP5C1& & & & & &0.0851 \\
rs4838643 &WDFY4&-0.0051 & & & & & \\
rs1916458 &RP11-170M17.1& & & & -0.0264& & \\
rs1537615 &RP11-526P5.2&0.0477 & & &  & & \\
rs2765398 &KRT8P37&-0.0157 & & &  & & \\
rs4317891 &CELF2&  & 0.0647& &  & & \\
rs7922793 &LINC00845&-0.0345  &  & &  & & \\
rs1916412 &RP11-170M17.1&  &  & & -0.0614 & & \\
rs1916411 &RP11-170M17.1&  &  & & -0.0448 & & \\
rs4747800 &KRT8P37&0.0036  &  & &  & & \\
rs11258040 &CAMK1D& &  & &  &-0.0983 & \\
rs1984275 &RP11-319F12.2& &  & &  &0.0065 & \\
rs17432763 &MIR5100& &  & &  & & -0.0677\\
rs10796113 &FRMD4A&-0.0931 &  & &  & &  \\
rs224765 &RP11-490O24.2&  & -0.0521 & &  & &  \\
rs6482387 &KIAA1217&  &  & &  & &0.011  \\
rs1492608 &ENKUR&  &  & &  & &-0.0287 \\
rs11257323 &ECHDC3&  &  & &  & &0.0084 \\
rs4434904 &KIAA1217&  &  & &  & &-0.0374 \\
rs10994364&ANK3&  &0.1086  & &  & &  \\
rs12220246&KIAA1462&  & & & 0.0371 & &  \\
rs11010390&RP11-309N24.1&  &0.0271 & &  & &  \\
rs10828584&KIAA1217&0.087  & & &  & &  \\
rs10857590&ARHGAP22& & & &-0.1379  & &  \\
rs1537616&RP11-526P5.2&0.086 & & & & &  \\
rs17295031&KIAA1462&  & & &-0.0468 & &  \\
rs10905778&RP11-271F18.4&  & & & & &0.0055  \\
rs7093161&SNRPEP8&  &-0.0577 & & & &0.0107 \\
rs2377872&CHAT&  &  & & &-0.0213 &  \\
rs1916409&RP11-170M17.1&  &  & & -0.0642 &  &  \\
rs2245456&MALRD1&  &-0.0042  & &  &  &  \\
rs787116&RP11-478H13.1&  &0.0259  & &  &  &  \\
rs2817825&RP11-492M23.2&  &0.0278  & &  &  &  \\
rs11255338&KIN&  &   & &  &  &-0.0401  \\
rs17011115&WDFY4&-0.0045  &   & &  &  &   \\
rs11010821&Y-RNA&-0.025  &   & &  &  &   \\
rs2532760&RP11-492M23.2& &0.0272  & &  &  &   \\
rs10821773&ANK3& &-0.0107  & &  &  &   \\
rs17454012&CELF2& &-0.1076 & &  &  &   \\
rs4372368&RP11-478B11.2& &-0.0449 & &  &  &   \\
rs1916420&RP11-170M17.1& &  & &0.0885  &  &   \\
rs2446588&FRMD4A& &-0.0142  & &  &  &   \\
rs10995687&RP11-170M17.1& & & & 0.1065 &  &   \\
rs161279&RP11-192P3.5& & & & 0.0333&  &   \\
rs161279&ZEB1& & & & 0.0333&  &   \\
rs161258&ZEB1& & & & 0.0362&  &   \\
rs10509149&TMEM26& & & &  &  &-0.0428   \\
rs3740000&LINC00837&-0.106 & & &  &  &  \\
rs17314489&ZNF365&  & 0.1543& &  &  &  \\
rs17453876&CELF2&  & 0.0518& &  &  &  \\
rs10793451&ZNF485&  &  & &-0.1028  &  &  \\
rs4749527&KIAA1462&  &  & &0.0093 &  &  \\
rs12570207&SEPHS1&  & -0.0329 & & &  &  \\
rs902904&THNSL1&  &  & & &  &-0.0885  \\
rs7921813&CAMK1D&  &0.0063  & & &  &  \\
rs10218945&SNRPEP8&  & & & &  &-0.0351  \\
rs2804551&RP11-492M23.2&0.0616  & & & &  &  \\
rs12266433&CELF2& &-0.0301 & & &  &  \\
rs16919385&PLXDC2& & & & &  &0.0112  \\
rs4750039&CELF2& &0.0333 & & &  &\\
rs12249964&KIAA1217& &  & & &  &0.0607 \\
rs4745829&RP11-170M17.1 &  & & &-0.0252  & \\
rs11257932&CAMK1D &  &0.0256 & & & \\
rs10827602&RP11-810B23.1 &  &  &0.0167 & & \\
rs7081466&RP11-526P5.2&  & &  & & 0.0267& \\
rs12256642&THNSL1&  &  & & & &-0.0258 \\
rs2796304&RP11-492M23.2&  &0.0875  & & &  \\
rs10826964&ZEB1&  & & &0.0316 & & \\
rs11257933&CAMK1D&  &0.0547 & &  & & \\
rs17432532&MIR5100&  &  & &  & &0.017 \\
rs10826899&UBE2V2P1&  &  & &  &-0.0234 & \\
rs11592473&UBE2V2P1&  &  & &  &-0.0642& \\
rs12764778&OR13A1&  &0.0263  & &  & & \\
rs12762462&GPR158&  &   & & -0.0153 & & \\
rs1011763&MIR3924&  &   & &  &0.1871 & \\
rs1916450&RP11-170M17.1&  &   & &-0.1767  &  & \\
rs1917814&CHAT&  &   &  &   & -0.0194 & \\
rs6602809&DCLRE1CP1&0.0043  &   &  &   &   & \\
rs923757&THNSL1&   &   &  &   &   &0.0226 \\
rs7092368&RP11-526P5.2&   &   &  &   & -0.0531  & \\
rs6602806&DCLRE1CP1&0.0527   &   &  &   &    & \\
rs6602806&ACBD7&0.0527   &   &  &   &    & \\
rs10994308&ANK3&   &   &  &   &    &-0.0124 \\
rs224699&RP11-490O24.2&   &-0.0351   &  &   &    &  \\
rs7083349&KIAA1217&-0.0651   &    &  &   &    &  \\
rs10828905&RNU6-632P&    &-0.0799    &  &   &    &  \\

\end{longtable}

\begin{longtable}{ c c c c c c c c c}

\caption[Inclusion probability of the NHS T2D data using LADBLSS.]{Inclusion probability of the NHS T2D data using LADBLSS.} \label{INCLU.LADBLSS} \\
\hline
SNP  &Gene      & Main Effects   & age   & act  & trans   & ceraf & chol \\
\hline	
	\endfirsthead
	\caption[]{Continued from the previous page.} \\ \hline

SNP  &Gene      & Main Effects   & age   & act  & trans   & ceraf & chol \\
\hline	

	\endhead
	\hline 
	\multicolumn{8}{r@{}}{Continued on the next page}\\
	\endfoot
	\hline
	\endlastfoot
rs17011106      & WDFY4 &0.9930  & & &&&\\
rs7077294 &KIAA1217 & &&&&&0.9736\\
rs7093682 &RP11-170M17.1& & & &0.9938 &&\\
rs17011106 &WDFY4&0.9900 & & & &&\\
rs10826028 &MIR3924&  & & & &0.9612&\\
rs4748996 &THNSL1&  & & & & &0.9834\\
rs2646392 &KRT8P37&0.9818  & & & & & \\
rs7904629 &RP11-170M17.1& & & &0.9646 & & \\
rs1244416 &ATP5C1& & & & & &0.9656 \\
rs4838643 &WDFY4&0.9768 & & & & & \\
rs1916458 &RP11-170M17.1& & & & 0.9832& & \\
rs1537615 &RP11-526P5.2&0.9956 & & &  & & \\
rs2765398 &KRT8P37&0.9756 & & &  & & \\
rs4317891 &CELF2&  & 1.000& &  & & \\
rs7922793 &LINC00845&0.9744  &  & &  & & \\
rs1916412 &RP11-170M17.1&  &  & & 0.9774 & & \\
rs1916411 &RP11-170M17.1&  &  & & 0.9700 & & \\
rs4747800 &KRT8P37&0.9840  &  & &  & & \\
rs11258040 &CAMK1D& &  & &  &0.9738 & \\
rs1984275 &RP11-319F12.2& &  & &  &0.9952 & \\
rs17432763 &MIR5100& &  & &  & & 0.9862\\
rs10796113 &FRMD4A&0.9636 &  & &  & &  \\
rs224765 &RP11-490O24.2&  &0.9710& &  & &  \\
rs6482387 &KIAA1217&  &  & &  & &0.9638  \\
rs1492608 &ENKUR&  &  & &  & &0.9680 \\
rs11257323 &ECHDC3&  &  & &  & &0.9892 \\
rs4434904 &KIAA1217&  &  & &  & &0.9716 \\
rs10994364&ANK3&  &0.9942  & &  & &  \\
rs12220246&KIAA1462&  & & & 0.9610 & &  \\
rs11010390&RP11-309N24.1&  &0.9820 & &  & &  \\
rs10828584&KIAA1217&0.9752  & & &  & &  \\
rs10857590&ARHGAP22& & & &0.9848  & &  \\
rs1537616&RP11-526P5.2&0.9944 & & & & &  \\
rs17295031&KIAA1462&  & & &0.9816 & &  \\
rs10905778&RP11-271F18.4&  & & & & &0.9988  \\
rs7093161&SNRPEP8&  &0.9542 & & & &0.9902 \\
rs2377872&CHAT&  &  & & &0.9728 &  \\
rs1916409&RP11-170M17.1&  &  & & 0.9612 &  &  \\
rs2245456&MALRD1&  &0.9630  & &  &  &  \\
rs787116&RP11-478H13.1&  &0.9638  & &  &  &  \\
rs2817825&RP11-492M23.2&  &0.9550  & &  &  &  \\
rs11255338&KIN&  &   & &  &  &0.9964  \\
rs17011115&WDFY4&0.9712  &   & &  &  &   \\
rs11010821&Y-RNA&0.9916  &   & &  &  &   \\
rs2532760&RP11-492M23.2& &0.9720  & &  &  &   \\
rs10821773&ANK3& &0.9586  & &  &  &   \\
rs17454012&CELF2& &0.9998 & &  &  &   \\
rs4372368&RP11-478B11.2& &0.9618 & &  &  &   \\
rs1916420&RP11-170M17.1& &  & &0.9672  &  &   \\
rs2446588&FRMD4A& &0.9724  & &  &  &   \\
rs10995687&RP11-170M17.1& & & & 0.9588 &  &   \\
rs161279&RP11-192P3.5& & & & 0.9770&  &   \\
rs161279&ZEB1& & & & 0.9770&  &   \\
rs161258&ZEB1& & & & 0.9876&  &   \\
rs10509149&TMEM26& & & &  &  &0.9726  \\
rs3740000&LINC00837&0.9964 & & &  &  &  \\
rs17314489&ZNF365&  & 0.9866& &  &  &  \\
rs17453876&CELF2&  & 0.9952& &  &  &  \\
rs10793451&ZNF485&  &  & &0.9604  &  &  \\
rs4749527&KIAA1462&  &  & &0.9794 &  &  \\
rs12570207&SEPHS1&  & 0.9698 & & &  &  \\
rs902904&THNSL1&  &  & & &  &0.9884  \\
rs7921813&CAMK1D&  &0.9998  & & &  &  \\
rs10218945&SNRPEP8&  & & & &  &0.9612  \\
rs2804551&RP11-492M23.2&0.9848 & & & &  &  \\
rs12266433&CELF2& &0.9618 & & &  &  \\
rs16919385&PLXDC2& & & & &  &0.9806 \\
rs4750039&CELF2& &0.9910 & & &  &\\
rs12249964&KIAA1217& &  & & &  &0.9558 \\
rs4745829&RP11-170M17.1 &  & & &0.9940  & \\
rs11257932&CAMK1D &  &0.9826 & & & \\
rs10827602&RP11-810B23.1 &  &  &0.9728 & & \\
rs7081466&RP11-526P5.2&  & &  & & 0.9714& \\
rs12256642&THNSL1&  &  & & & &0.9616 \\
rs2796304&RP11-492M23.2&  &0.9928  & & &  \\
rs10826964&ZEB1&  & & &0.9592& & \\
rs11257933&CAMK1D&  &0.9726 & &  & & \\
rs17432532&MIR5100&  &  & &  & &0.9834 \\
rs10826899&UBE2V2P1&  &  & &  &0.9784 & \\
rs11592473&UBE2V2P1&  &  & &  &0.9864& \\
rs12764778&OR13A1&  &0.9894  & &  & & \\
rs12762462&GPR158&  &   & & 0.9636 & & \\
rs1011763&MIR3924&  &   & &  &0.9954 & \\
rs1916450&RP11-170M17.1&  &   & &0.9820 &  & \\
rs1917814&CHAT&  &   &  &   & 0.9670 & \\
rs6602809&DCLRE1CP1&0.9614  &   &  &   &   & \\
rs923757&THNSL1&   &   &  &   &   &0.9964 \\
rs7092368&RP11-526P5.2&   &   &  &   & 0.9868  & \\
rs6602806&DCLRE1CP1&0.9912   &   &  &   &    & \\
rs6602806&ACBD7&0.9912  &   &  &   &    & \\
rs10994308&ANK3&   &   &  &   &    &0.9542 \\
rs224699&RP11-490O24.2&   &0.9768   &  &   &    &  \\
rs7083349&KIAA1217&0.9986   &    &  &   &    &  \\
rs10828905&RNU6-632P&    &0.9626    &  &   &    &  \\

\end{longtable}

\clearpage
\section{Posterior inference}

\subsection{LADBL}
\subsubsection{Hierarchical model specification}

\begin{gather*}
Y_{i} = E_{i}\alpha + C_{i}\gamma + X_{ij}\beta_{j} + \tilde{W}_{i}\eta_{j} + \tau^{-1/2}\xi_{2}\sqrt{v_{i}}z_{i} \quad i = 1, \dots,n\\
v_{i}|\tau \stackrel{iid}{\thicksim}\tau \text{exp}(- \tau v_{i}) \quad i = 1, \dots,n\\
z_{i} \stackrel{iid}{\thicksim}\text{N}(0,1) \quad i = 1, \dots,n\\
\beta_j|s_{1}{\thicksim}\frac{1}{\sqrt{2\pi s_{1}}}\text{exp}(-\frac{\beta_{j}^{2}}{2 s_{1}})\\
s_{1}|{\varphi_{1}^{2}}{\thicksim}\frac{\varphi_{1}^{2}}{2}\text{exp}({-\frac{\varphi_{1}^{2}}{2}}s_{1})\\
\eta_{jk}|s_{2k} \stackrel{iid}{\thicksim}\frac{1}{\sqrt{2\pi s_{2k}}}\text{exp}(-\frac{\eta_{jk}^2}{2 s_{2k}}) \quad k=1,\dots, q\\
s_{2k}|{\varphi_{2}^{2}}\stackrel{iid}{\thicksim}\frac{\varphi_{2}^{2}}{2}\text{exp}({-\frac{\varphi_{2}^{2}}{2}}s_{2k}) \quad k=1,\dots, q\\
\alpha_k \stackrel{iid}{\thicksim}\frac{1}{\sqrt{(2\pi \alpha_0)}}\text{exp}(-\frac{\alpha_{k}^2}{2\alpha_0})\quad k = 1,\dots, q\\
\gamma_t \stackrel{iid}{\thicksim}\frac{1}{\sqrt{(2\pi \gamma_0)}}\text{exp}(-\frac{\gamma_{t}^2}{2\gamma_0}) \quad t = 1, \dots, m\\
\tau {\thicksim}\text{Gamma}(a,b)\\
\varphi_1^2{\thicksim}\text{Gamma}(c_1,d_1)\\
\varphi_2^2{\thicksim}\text{Gamma}(c_2,d_2)\\
\end{gather*}

\subsubsection{Gibbs Sampler}

Let $\mu_{(-\alpha_k)}=E(y_i)-E_{ik}\alpha_k$, then
\begin{equation*}
\begin{aligned}
\pi(\alpha_k|& \text{rest}) \\
& \propto \pi(Y|\cdot) \pi(\alpha_k) \\
& \propto \exp\Big\{- \sum_{i=1}^{n}\frac{(y_i - E_i \alpha - C_i \gamma - X_{ij}\beta_{j} - \tilde{W}_{i}\eta_{j})^2}{2 \tau^{-1} \xi_{2}^{2} v_i} \Big\}
 \times\exp(-\frac{\alpha_{k}^2}{2 \alpha_0})\\
& \propto \exp\Big\{ -\frac{1}{2}\big[(\sum_{i=1}^{n} \frac{\tau E_{ik}^{2}}{\xi_{2}^{2} v_i}+\frac{1}{\alpha_0})\alpha_k^2 - 2\sum_{i=1}^{n}\frac{\tau(y_i-\mu_{(-\alpha_k)})E_{ik}}{\xi_2^2 v_i}\alpha_k\big]\Big\}.
\end{aligned}
\end{equation*}

Hence, $\alpha_k| \text{rest} {\thicksim} N(\mu_{\alpha_k}, \sigma_{\alpha_k}^{2})$, where
\begin{equation*}
\begin{aligned}
\mu_{\alpha_k} &= \big(\sum_{i=1}^{n}\frac{\tau(y_i-\mu_{(-\alpha_k)})E_{ik}}{\xi_2^2 v_i}\big)\sigma_{\alpha_k}^2,\\
\sigma_{\alpha_k}^{2} &= \big( \sum_{i=1}^{n} \frac{\tau E_{ik}^2}{\xi_2^2 v_i}+\frac{1}{\alpha_0}\big)^{-1}.
\end{aligned}
\end{equation*}

Let $\mu_{(-\gamma_t)}=E(y_i)-C_{it}\gamma_t\ $, So $\gamma_t| \text{rest} {\thicksim} N(\mu_{\gamma_k}, \sigma_{\gamma_t}^{2})$,
where 
\begin{equation*}\label{equr:meanm}
\begin{aligned}
\mu_{\gamma_t} &= \big(\sum_{i=1}^{n}\frac{\tau(y_i-\mu_{(-\gamma_t)})C_{it}}{\xi_2^2 v_i}\big)\sigma_{\gamma_t}^2,\\
\sigma_{\gamma_t}^{2} &= \big( \sum_{i=1}^{n} \frac{\tau C_{it}^2}{\xi_2^2 v_i}+\frac{1}{\gamma_0}\big)^{-1}.
\end{aligned}
\end{equation*}

Let $\mu_{(-\beta_j)}=E(y_i)-X_{ij}\beta_j $, then
\begin{equation*}
\begin{aligned}
\pi(\beta_j|& \text{rest}) \\
& \propto \pi(y|\cdot) \pi(\beta_j|s_1) \\
& \propto \exp\Big\{- \sum_{i=1}^{n}\frac{(y_i - E_i \alpha - C_i \gamma - X_{ij}\beta_{j} - \tilde{W}_{i}\eta_{j})^2}{2 \tau^{-1} \xi_{2}^{2} v_i} \Big\}
 \times\exp(-\frac{\beta_j^2}{2s_1})\\
& \propto \exp\Big\{ -\frac{1}{2}\big[(\sum_{i=1}^{n} \frac{\tau X_{ij}^{2}}{\xi_{2}^{2} v_i}+\frac{1}{s_1})\beta_j^2 - 2\sum_{i=1}^{n}\frac{\tau(y_i-\mu_{(-\beta_j)})X_{ij}}{\xi_2^2 v_i}\beta_j\big]\Big\}.
\end{aligned}
\end{equation*}

So, $\beta_j|\text{rest}{\thicksim} N(\mu_{\beta_j}, \sigma_{\beta_j}^2)$ with
\begin{equation*}
\begin{aligned}
\mu_{\beta_j} &= \big(\sum_{i=1}^{n}\frac{\tau(y_i-\mu_{(-\beta_j)})X_{ij}}{\xi_2^2 v_i}\big)\sigma_{\beta_j}^2,\\
\sigma_{\beta_j}^{2} &= \big( \sum_{i=1}^{n} \frac{\tau X_{ij}^2}{\xi_2^2 v_i}+\frac{1}{s_1}\big)^{-1}.
\end{aligned}
\end{equation*}

Let $\mu_{(-\eta_{jk})}=E(y_i)-W_{ik}\eta_{jk}$, then $\eta_{jk}|\text{rest}{\thicksim}  N(\mu_{\eta_{jk}}, \sigma_{\eta_{jk}}^2)$, where 
\begin{equation*}
\begin{aligned}
\mu_{\eta_{jk}} &= \big(\sum_{i=1}^{n}\frac{\tau(y_i-\mu_{(-\eta_{jk})})\tilde{W}_{ik}}{\xi_2^2 v_i}\big)\sigma_{\eta_{jk}}^2,\\
\sigma_{\beta_j}^{2} &= \big( \sum_{i=1}^{n} \frac{\tau \tilde{W}_{ik}^2}{\xi_2^2 v_i}+\frac{1}{s_{2k}}\big)^{-1}.
\end{aligned}
\end{equation*}

The full conditional posterior distribution of $s_1$ is:
\begin{equation*}
\begin{aligned}
s_{1}&|\text{rest} \\
&\propto \pi(\beta_{j}|s_1) \pi(s_1|\varphi_{1}^{2})\\
&\propto \frac{1}{\sqrt{s_1}} \text{exp}(-\frac{\varphi_{1}^{2}}{2} s_1) \text{exp}(-\frac{\beta_{j}^{2}}{2 s_1})\\
&\propto \frac{1}{\sqrt{s_1}} \text{exp} \Big\{ -\frac{1}{2}[\varphi_{1}^{2} s_1 + \frac{\beta_{j}^{2}}{s_1} ]  \}.\\
\end{aligned}
\end{equation*}
Therefore, $s_1^{-1}|\text{rest}{\thicksim} \text{ Inverse-Gaussian}(\sqrt{\frac{\varphi_1^2}{\beta_j^2}}, \varphi_1^2)$.\\

Similiarly, for $s_{2k} (k=1,\dots,q)$, the posterior distribution for is $s_{2k}^{-1}|\text{rest}{\thicksim} \text{ Inverse-Gaussian}(\sqrt{\frac{\varphi_2^2}{\eta_{jk}^2}}, \varphi_2^2)$.\par

The full conditional posterior distribution of $\varphi_1^2$ is:
\begin{equation*}
\begin{aligned}
\varphi_1^2&|\text{rest} \\
&\propto \pi(s_1|\varphi_1^2)\pi(\varphi_1^2)\\
&\propto \frac{\varphi_1^2}{2} \text{exp}(-\frac{\varphi_1^2 s_1}{2})(\varphi_1^2)^{c_1-1}\text{exp}(-d_1\varphi_1^2)\\
&\propto (\varphi_1^2)^{c_1} \text{exp}\Big(- \varphi_1^2(s_1/2+d_1)\Big).
\end{aligned}
\end{equation*}
Therefore, the posterior distribution for $\varphi_1^2$ is Gamma($c_1+1, s_1/2+d_1$).\par

The full conditional posterior distribution of $\varphi_2^2$ is:
\begin{equation*}
\begin{aligned}
\varphi_2^2&|\text{rest} \\
&\propto \pi(s_2|\varphi_2^2)\pi(\varphi_2^2)\\
&\propto \prod_{k=1}^{q}\frac{\varphi_2^2}{2} \text{exp}(-\frac{\varphi_2^2 s_{2k}}{2})(\varphi_2^2)^{c_2-1}\text{exp}(-d_2\varphi_2^2)\\
&\propto (\varphi_2^2)^{q+c_2-1} \text{exp}\Big(- \varphi_2^2(\sum_{k=1}^{q}\frac{s_{2k}}{2}+d_2)\Big).
\end{aligned}
\end{equation*}
The posterior distribution for $\varphi_2^2$ is Gamma($c_2+q, \sum_{k=1}^{q} s_{2k}/2+d_2$).\par

The full conditional posterior distribution of $\tau$:
\begin{equation*}
\begin{aligned}
\tau&|\text{rest} \\
& \propto \pi(v|\tau)\pi(\tau)\pi(Y|\cdot) \\
& \propto {\tau}^{n/2}\exp\Big\{- \sum_{i=1}^{n}\frac{(y_i - E_i \alpha - C_i \gamma - X_{ij}\beta_{j} - \tilde{W}_{i}\eta_{j})^2}{2 \tau^{-1} \xi_{2}^{2} v_i} \Big\}\\
& \times {\tau}^n \text{exp}(-\tau\sum_{i=1}^{n} v_i) \tau^{a-1} \text{exp}(-b\tau)\\
&\propto \tau^{a+\frac{3}{2}n-1} \text{exp}\Big\{-\tau\big[\sum_{i=1}^{n}(\frac{(y_i - E_i \alpha - C_i \gamma - X_{ij}\beta_{j} - \tilde{W}_{i}\eta_{j})^2}{2  \xi_{2}^{2} v_i} +v_i)+b\big] \Big\}.
\end{aligned}
\end{equation*}
Therefore, $\tau|\text{rest} {\thicksim} \text{Gamma}(a+\frac{3}{2}n, \:\big[\sum_{i=1}^{n}(\frac{(y_i - E_i \alpha - C_i \gamma - X_{ij}\beta_{j} - \tilde{W}_{i}\eta_{j})^2}{2  \xi_{2}^{2} v_i} +v_i)+b\big])$.\par
The full conditional posterior distribution of $v_i$ is:
\begin{equation*}
\begin{aligned}
v_i&|\text{rest} \\
& \propto \pi(v|\tau)\pi(y|\cdot) \\
& \propto \frac{1}{\sqrt{v_i}} \exp\Big\{- \frac{(y_i - E_i \alpha - C_i \gamma - X_{ij}\beta_{j} - \tilde{W}_{i}\eta_{j})^2}{2 \tau^{-1} \xi_{2}^{2} v_i} \Big\}
\times\exp(-\tau v_i)\\
& \propto \frac{1}{\sqrt{v_i}} \exp\Big\{-\frac{1}{2} \big[ (2\tau)v_i + \frac{\tau(y_i - E_i \alpha - C_i \gamma - X_{ij}\beta_{j} - \tilde{W}_{i}\eta_{j})^2}{\xi_{2}^{2} v_i}\big] \Big\}.
\end{aligned}
\end{equation*}
Therefore,
\begin{equation*}
\frac{1}{v_i}|\text{rest} {\thicksim}\text{ Inverse-Gaussian} (\sqrt{\frac{2 \xi_2^2}{(y_i - E_i \alpha - C_i \gamma - X_{ij}\beta_{j} - \tilde{W}_{i}\eta_{j})^2}}, \:2\tau).
\end{equation*}

\subsection{BLSS}
\subsubsection{Hierarchical model specification}
\begin{gather*}
Y \propto (\sigma^{2})^{-\frac{n}{2}} \exp\left\{-\frac{1}{2\sigma^{2}} \sum_{i=1}^{n}(y_{i}- E_{i}\alpha - C_{i}\gamma - X_{ij}\beta_{j} - \tilde{W}_{i}\eta_{j})^{2} \right\} \\ 
\alpha \thicksim \text{N}_{q}(0, \, \Sigma_{\alpha0}) \\
\gamma \thicksim \text{N}_{m}(0, \, \Sigma_{\gamma0}) \\
\beta_{j}|\pi_{c}, \tau_{c}^{2}, \sigma^{2} {\thicksim} (1-\pi_{c})\, \text{N} \left(0,\, \sigma^{2} \tau_{c}^{2}\right) + \pi_c \,\delta_{0}(\beta_{j}) \quad j=1,\dots,p\\
\eta_{jk}|\pi_{e}, \tau_{ek}^{2}, \sigma^{2} \stackrel{iid}{\thicksim} (1-\pi_{e})\, \text{N} \left(0,\, \sigma^{2} \tau_{ek}^{2}\right) + \pi_e \,\delta_{0}(\eta_{jk}) \quad j=1,\dots,p,\, k=1,\dots,q \\
\tau_{c}^{2}|\lambda_{c}^{2}{\thicksim} \text{Gamma} (1, \frac{\lambda_{c}^{2}}{2})\\
\tau_{ek}^{2}|\lambda_{e}^{2}\stackrel{iid} {\thicksim} \text{Gamma} (1, \frac{\lambda_{e}^{2}}{2})\quad k=1,\dots,q\\
\pi_{c} \thicksim \text{Beta}\left(r_{c}, \, u_{c}\right) \\
\pi_{e} \thicksim \text{Beta}\left(r_{e}, \, u_{e}\right) \\
\lambda_{c}^2 \thicksim \text{Gamma}\left(a_{c}, \, b_{c}\right) \\
\lambda_{e}^2 \thicksim \text{Gamma}\left(a_{e}, \, b_{e}\right) \\
\sigma^{2} \thicksim \text{Inverse-Gamma}\left(s, \, h\right) 
\end{gather*}

\subsubsection{Gibbs Sampler}
Denote $\mu_{(-\alpha)}=\text{E}(Y)-E\alpha$, then $\alpha|\text{rest} \thicksim \text{N}(\mu_{\alpha}, \, \Sigma_{\alpha})$, where
\begin{equation*}
	\begin{aligned}
	\mu_{\alpha} &= \Sigma_{\alpha}(\frac{1}{\sigma^2}(Y-\mu_{(-\alpha)})^\top E)^\top, \\
	\Sigma_{\alpha} &= \left(\frac{1}{\sigma^{2}}E^\top E + \Sigma_{\alpha0}^{-1} \right)^{-1} .
	\end{aligned}
	\end{equation*}
Denote $\mu_{(-\gamma)}=\text{E}(Y)-C\gamma$, then $\gamma|\text{rest} \thicksim \text{N}(\mu_{\gamma}, \, \Sigma_{\gamma})$, where
\begin{equation*}
	\begin{aligned}
	\mu_{\gamma} &= \Sigma_{\gamma}(\frac{1}{\sigma^2}(Y-\mu_{(-\gamma)})^\top C)^\top, \\
	\Sigma_{\gamma} &= \left(\frac{1}{\sigma^{2}}C^\top C + \Sigma_{\gamma0}^{-1} \right)^{-1}. 
	\end{aligned}
	\end{equation*}

Denote $\mu_{(-\beta_j)}=\text{E}(Y)-X_{j}\beta_j$, then $\beta_{j}|\text{rest} \thicksim (1-l_{c}) \text{N}(\mu_{\beta_{j}}, \, \sigma^{2} \Sigma_{\beta_{j}}) + l_{c}\delta_{0}(\beta_{j})$, where
	\begin{equation*}
	\begin{aligned}
	\mu_{\beta_{j}} &= \Sigma_{\beta_{j}} X_{j}^\top(Y-\mu_{(-\beta_j)}), \\
	\Sigma_{\beta_{j}} &= \left(X_{j}^\top X_{j} + \frac{1}{\tau_{c}^{2}} \right)^{-1},\\
	l_{c} &= \frac{\pi_{c}}{\pi_{c} + (1-\pi_{c})(\tau_{c}^{2})^{-1/2}{|\Sigma_{\beta_{j}}|}^{1/2}\text{exp}\big\{ \frac{1}{2 \sigma^2} \Sigma_{\beta_{j}}{||X_{j}^\top(Y-\mu_{(-\beta_j)}||}_{2}^{2}\big\} }.
	\end{aligned}
	\end{equation*}

Denote $\mu_{(-\eta_{jk})}=\text{E}(Y)-\tilde{W}_{k}\eta_{jk}$, then $\eta_{jk}|\text{rest} \thicksim (1-l_{ek}) \text{N}(\mu_{\eta_{jk}}, \, \sigma^{2}\Sigma_{\eta_{jk}}) + l_{e}\delta_{0}(\eta_{jk})$, where
	\begin{equation*}
	\begin{aligned}
	\mu_{\eta_{jk}} &= \Sigma_{\eta_{jk}} {\tilde{W}_{k}}^\top(Y-\mu_{(-\eta_{jk})}), \\
	\Sigma_{\eta_{jk}} &= \left(\tilde{W}_{k}^\top \tilde{W}_{k} + \frac{1}{\tau_{ek}^{2}} \right)^{-1},\\
	l_{e} &= \frac{\pi_{e}}{\pi_{e} + (1-\pi_{e})(\tau_{ek}^{2})^{-1/2}{|\Sigma_{\eta_{jk}}|}^{1/2}\text{exp}\big\{ \frac{1}{2 \sigma^2} \Sigma_{\eta_{jk}}{||\tilde{W}_{k}^\top(Y-\mu_{(-\eta_{jk})})||}_2^{2}\big\} }.
	\end{aligned}
	\end{equation*}

The posterior of $\tau_{c}^{2}$ is:
	\begin{equation*}
	\frac{1}{\tau_{c}^{2}}|\text{rest} \thicksim \begin{cases}
	\scalebox{1}{Inverse-Gamma($1$,\,$\frac{\lambda_{c}^{2}}{2}$)}& { \text{if} \; \beta_{j} = 0}\\
	\scalebox{1}{Inverse-Gaussian($\sqrt{\frac{\sigma^{2}}{\beta_{j}^{2}}\lambda_{c}^{2}}$, $\lambda_{c}^{2}$)}& { \text{if} \; \beta_{j} \neq 0}
	\end{cases}.
	\end{equation*}

The posterior of $\tau_{ek}^{2}$ is:
	\begin{equation*}
	\frac{1}{\tau_{ek}^{2}}|\text{rest} \thicksim \begin{cases}
	\scalebox{1}{Inverse-Gamma($1$,\,$\frac{\lambda_{e}^{2}}{2}$)}& { \text{if} \; \eta_{jk} = 0} \\
	\scalebox{1}{Inverse-Gaussian($\sqrt{\frac{\sigma^{2}}{\eta_{jk}^{2}}\lambda_{e}^{2}}$, $\lambda_{e}^{2}$)}& { \text{if} \; \eta_{jk} \neq 0}
	\end{cases}.
	\end{equation*}

$\lambda_{c}^{2}$ and $\lambda_{e}^{2}$ have Gamma posterior distributions:
\begin{equation*}
\begin{aligned}
\lambda_{c}^{2}|\text{rest} & \thicksim \text{Gamma} (a_c+1,\; \frac{\tau_{c}^{2}}{2}+b_c),\\
\lambda_{e}^{2}|\text{rest} & \thicksim \text{Gamma} (a_e+q,\; \sum_{k=1}^{q}\frac{\tau_{ek}^{2}}{2}+b_e).\\
\end{aligned}
\end{equation*}

$\pi_{c}$ and $\pi_{e}$ have Gamma posterior distributions:
\begin{equation*}
\begin{aligned}
\pi_{c}|\text{rest}& \thicksim \text{Beta}(r_{c}-\textbf{I}_{\{\beta_{j} \neq 0\}}+1, \; u_{c}+ \textbf{I}_{\{\beta_{j} \neq 0\}}),\\
\pi_{e}|\text{rest} &\thicksim \text{Beta}(r_{e}-\sum_{k=1}^{q}\textbf{I}_{\{\eta_{jk} \neq 0\}}+q, \; u_{e}+\sum_{k=1}^{q}\textbf{I}_{\{\eta_{jk} \neq 0\}}).
\end{aligned}
\end{equation*}

$\sigma^2 \thicksim \text{Inverse-Gamma}(\mu_{\sigma^2}, \; \Sigma_{\sigma^2})$, where
\begin{equation*}
\begin{aligned}
\mu_{\sigma^2} &= s+\frac{n+\textbf{I}_{\{\beta_{j} \neq 0\}}+\sum_{k=1}^{q}\textbf{I}_{\{\eta_{jk} \neq 0\}}}{2},\\
\Sigma_{\sigma^2} &= h+\frac{(Y-\mu)^\top (Y-\mu)+(\tau_{c}^{2})^{-1}\beta_{j}^{2} + \sum_{k=1}^{q}(\tau_{ek}^{2})^{-1} \eta_{j}^\top \eta_{j}}{2}.
\end{aligned}
\end{equation*}

\subsection{BL}
\subsubsection{Hierarchical model specification}

\begin{gather*}
Y \propto (\sigma^{2})^{-\frac{n}{2}} \exp\left\{-\frac{1}{2\sigma^{2}} \sum_{i=1}^{n}(y_{i}- E_{i}\alpha - C_{i}\gamma - X_{ij}\beta_{j} - \tilde{W}_{i}\eta_{j})^{2} \right\} \\ 
\alpha \thicksim \text{N}_{q}(0, \, \Sigma_{\alpha0}) \\
\gamma \thicksim \text{N}_{m}(0, \, \Sigma_{\gamma0}) \\
\beta_{j}| \tau_{c}^{2}, \sigma^{2} {\thicksim} \text{N} \left(0,\, \sigma^{2} \tau_{c}^{2}\right)  \quad j=1,\dots,p\\
\eta_{jk}| \tau_{ek}^{2}, \sigma^{2} \stackrel{iid}{\thicksim} \text{N} \left(0,\, \sigma^{2} \tau_{ek}^{2}\right) \quad j=1,\dots,p,\, k=1,\dots,q \\
\tau_{c}^{2}|\lambda_{c}^{2}{\thicksim} \text{exp} (\frac{\lambda_{c}^{2}}{2})\\
\tau_{ek}^{2}|\lambda_{e}^{2}\stackrel{iid} {\thicksim} \text{exp} (\frac{\lambda_{e}^{2}}{2})\quad k=1,\dots,q\\
\lambda_{c}^2 \thicksim \text{Gamma}\left(a_{c}, \, b_{c}\right) \\
\lambda_{e}^2 \thicksim \text{Gamma}\left(a_{e}, \, b_{e}\right) \\
\sigma^{2} \propto \frac{1}{\sigma^2}
\end{gather*}

\subsubsection{Gibbs Sampler}
Denote $\mu_{(-\alpha)}=\text{E}(Y)-E\alpha$, then $\alpha|\text{rest} \thicksim \text{N}(\mu_{\alpha}, \, \Sigma_{\alpha})$, where
\begin{equation*}
	\begin{aligned}
	\mu_{\alpha} &= \Sigma_{\alpha}(\frac{1}{\sigma^2}(Y-\mu_{(-\alpha)})^\top E)^\top, \\
	\Sigma_{\alpha} &= \left(\frac{1}{\sigma^{2}}E^\top E + \Sigma_{\alpha0}^{-1} \right)^{-1}.
	\end{aligned}
	\end{equation*}
Denote $\mu_{(-\gamma)}=\text{E}(Y)-C\gamma$, then $\gamma|\text{rest} \thicksim \text{N}(\mu_{\gamma}, \, \Sigma_{\gamma})$, where
\begin{equation*}
	\begin{aligned}
	\mu_{\gamma} &= \Sigma_{\gamma}(\frac{1}{\sigma^2}(Y-\mu_{(-\gamma)})^\top C)^\top, \\
	\Sigma_{\gamma} &= \left(\frac{1}{\sigma^{2}}C^\top C + \Sigma_{\gamma0}^{-1} \right)^{-1}. 
	\end{aligned}
	\end{equation*}

Denote $\mu_{(-\beta_j)}=\text{E}(Y)-X_{j}\beta_j$, then $\beta_{j}|\text{rest} \thicksim \text{N}(\mu_{\beta_{j}}, \, \sigma^{2} \Sigma_{\beta_{j}})$, where
	\begin{equation*}
	\begin{aligned}
	\mu_{\beta_{j}} &= \Sigma_{\beta_{j}} X_{j}^\top(Y-\mu_{(-\beta_j)}), \\
	\Sigma_{\beta_{j}} &= \left(X_{j}^\top X_{j} + \frac{1}{\tau_{c}^{2}} \right)^{-1}. \\
	\end{aligned}
	\end{equation*}

Denote $\mu_{(-\eta_{jk})}=\text{E}(Y)-\tilde{W}_{k}\eta_{jk}$, then $\eta_{jk}|\text{rest} \thicksim \text{N}(\mu_{\eta_{jk}}, \, \sigma^{2}\Sigma_{\eta_{jk}})$, where
	\begin{equation*}
	\begin{aligned}
	\mu_{\eta_{jk}} &= \Sigma_{\eta_{jk}} {\tilde{W}_{k}}^\top(Y-\mu_{(-\eta_{jk})}), \\
	\Sigma_{\eta_{jk}} &= \left(\tilde{W}_{k}^\top \tilde{W}_{k} + \frac{1}{\tau_{ek}^{2}} \right)^{-1}. \\
	\end{aligned}
	\end{equation*}

The posterior of $\tau_{c}^{2}$ is:
	\begin{equation*}
	\frac{1}{\tau_{c}^{2}}|\text{rest} \thicksim \text{Inverse-Gaussian}(\sqrt{\frac{\sigma^{2}}{\beta_{j}^{2}}\lambda_{c}^{2}}, \lambda_{c}^{2}).
	\end{equation*}

The posterior of $\tau_{ek}^{2}$ is:
	\begin{equation*}
	\frac{1}{\tau_{ek}^{2}}|\text{rest} \thicksim \text{Inverse-Gaussian}(\sqrt{\frac{\sigma^{2}}{\eta_{jk}^{2}}\lambda_{e}^{2}}, \lambda_{e}^{2}).
	\end{equation*}

$\lambda_{c}^{2}$ and $\lambda_{e}^{2}$ have Gamma posterior distributions:
\begin{equation*}
\begin{aligned}
\lambda_{c}^{2}|\text{rest} & \thicksim \text{Gamma} (a_c+1,\; \frac{\tau_{c}^{2}}{2}+b_c),\\
\lambda_{e}^{2}|\text{rest} & \thicksim \text{Gamma} (a_e+q,\; \sum_{k=1}^{q}\frac{\tau_{ek}^{2}}{2}+b_e).\\
\end{aligned}
\end{equation*}

$\sigma^2 \thicksim \text{Inverse-Gamma}(\mu_{\sigma^2}, \; \Sigma_{\sigma^2})$, where
\begin{equation*}
\begin{aligned}
\mu_{\sigma^2} &= \frac{n+1+q}{2},\\
\Sigma_{\sigma^2} &= \frac{(Y-\mu)^\top (Y-\mu)+(\tau_{c}^{2})^{-1}\beta_{j}^{2} + \sum_{k=1}^{q}(\tau_{ek}^{2})^{-1} \eta_{j}^\top \eta_{j}}{2}.
\end{aligned}
\end{equation*}


\begin{thebibliography}{99}
	
\bibitem[1]{HUNTER}Hunter, D. J. (2005). Gene–environment interactions in human diseases. \emph{Nature Reviews Genetics}, 6(4), 287-298.

\bibitem[2]{SIM}Simonds, N. I., Ghazarian, A. A., Pimentel, C. B., Schully, S. D., Ellison, G. L., Gillanders, E. M. and Mechanic, L. E. (2016). Review of the gene‐environment interaction literature in cancer: what do we know?. \emph{Genetic epidemiology}, 40(5), 356-365.

\bibitem[3]{VON}Von Mutius, E. (2009). Gene-environment interactions in asthma. \emph{Journal of Allergy and Clinical Immunology}, 123(1), 3-11.

\bibitem[4]{CORN}Cornelis, M. C. and Hu, F. B. (2012). Gene-environment interactions in the development of type 2 diabetes: recent progress and continuing challenges. \emph{Annual review of nutrition}, 32, 245-259.

\bibitem[5]{CORD}Cordell, H. J. and Clayton, D. G. (2005). Genetic association studies. \emph{The Lancet}, 366(9491), 1121-1131.

\bibitem[6]{WLC}Wu, C., Li, S. and Cui, Y. (2012). Genetic association studies: an information content perspective. \emph{Current genomics}, 13(7), 566-573.

\bibitem[7]{ZFR}Zhou, F., Ren, J., Lu, X., Ma, S. and Wu, C.  (2021). Gene–Environment Interaction: a Variable Selection Perspective. Epistasis. \emph{Methods in Molecular Biology}. Humana Press.

\bibitem[8]{SLH}Shi, X., Liu, J., Huang, J., Zhou, Y., Xie, Y. and Ma, S. (2014). A penalized robust method for identifying gene–environment interactions. \emph{Genetic epidemiology}, 38(3), 220-230.

\bibitem[9]{CZJ}Chai, H., Zhang, Q., Jiang, Y., Wang, G., Zhang, S., Ahmed, S. E. and Ma, S. (2017). Identifying gene-environment interactions for prognosis using a robust approach. \emph{Econometrics and statistics}, 4, 105-120.

\bibitem[10]{ZSX}Zhang, S., Xue, Y., Zhang, Q., Ma, C., Wu, M. and Ma, S. (2020). Identification of gene–environment interactions with marginal penalization. \emph{Genetic epidemiology}, 44(2), 159-196.

\bibitem[11]{HJM}Huang, J. and Ma, S. (2010). Variable selection in the accelerated failure time model via the bridge method. \emph{Lifetime data analysis}, 16(2), 176-195.

\bibitem[12]{LMA}Liu, C., Ma, J. and Amos, C. I. (2015). Bayesian variable selection for hierarchical gene–environment and gene–gene interactions. \emph{Human genetics}, 134(1), 23-36.

\bibitem[13]{LWL}Li, J., Wang, Z., Li, R. and Wu, R. (2015). Bayesian group lasso for nonparametric varying-coefficient models with application to functional genome-wide association studies. \emph{The annals of applied statistics}, 9(2), 640.

\bibitem[14]{RZL}Ren, J., Zhou, F., Li, X., Chen, Q., Zhang, H., Ma, S., Jiang, Y. and Wu, C. (2020). Semiparametric Bayesian variable selection for gene‐environment interactions.
 \emph{Statistics in Medicine}, 39(5), 617-638.

\bibitem[15]{GEMC}George, E. I. and McCulloch, R. E. (1993). Variable selection via Gibbs sampling. \emph{Journal of the American Statistical Association}, 88(423), 881-889.

\bibitem[16]{ISHW}Ishwaran, H. and Rao, J. S. (2005). Spike and slab variable selection: frequentist and Bayesian strategies. \emph{The Annals of Statistics}, 33(2), 730-773.

\bibitem[17]{WUM} Wu, C. and Ma, S. (2015) A selective review of robust variable selection with applications in bioinformatics, \emph{Briefings in Bioinformatics}, 16(5): 873-–883, 

\bibitem[18]{KOZU} Kozumi, H. and Kobayashi, G. (2011) Gibbs sampling methods for bayesian quantile regression.  \emph{Journal of Statistical Computation and Simulation}, 81(11):1565–1578. 

\bibitem[19]{MOY}Yu, K. and Moyeed, R.A. (2001) Bayesian quantile regression.  \emph{Statistics and Probability Letters}, 54(4): 437–447.

\bibitem[20]{YUK} Yu, K. and Zhang, J. (2005)  A three-parameter asymmetric laplace distribution and its extension.   \emph{Communications in Statistics - Theory and Methods}, 34(9-10):1867–-1879

\bibitem[21]{NANLIN}Li, Q., Xi, R., and Lin, N. (2010) Bayesian regularized quantile regression.   \emph{Bayesian Analysis}, 5(3):533–-556.

\bibitem[22]{BROOK}Brooks, S.P. and Gelman, A. (1998) General methods for monitoring convergence of iterative simulations.  \emph{Journal of Computational and Graphical Statistics}, 7(4):434–-455.

\bibitem[23]{GEL} Gelman, A., Carlin, J.B., Stern, H.S., Dunson, D.B., Vehtari, A. and Rubin, D.B. \emph{Bayesian Data Analysis}. Chapman and Hall/CRC, 2004.

\bibitem[24]{BAR} Barclay, S.F., Rand, C.M., Borch, L.A., Nguyen, L., Gray, P.A., Gibson, W.T., Wilson, R.J., Gordon, P.M., Aung, Z., Berry-Kravis, E.M., Ize-Ludlow, D., Weese-Mayer, D.E. and  Bech-Hansen, N.T. (2015) Rapid-Onset Obesity with Hypothalamic Dysfunction, Hypoventilation, and Autonomic Dysregulation (ROHHAD): exome sequencing of trios, monozygotic twins and tumours.  \emph{Orphanet Journal of Rare Diseases}, 10:103.

\bibitem[25]{MARTIN} Martin, C.L., Jima, D., Sharp, G.C., McCullough, L.E., Park, S.S., Gowdy, K.M., Skaar, D., Cowley, M., Maguire, R.L., Fuemmeler, B., Collier, D., Relton, C.L., Murphy, S.K., and Hoyo, C. (2019) Maternal pre-pregnancy obesity, offspring cord blood DNA methylation, and offspring cardiometabolic health in early childhood: an epigenome-wide association study. \emph{Epigenetics}, 14:4, 325-340. 

\bibitem[26]{GHAN} Ghanbari, M., de Vries, P.S., de Looper, H., Peters, M.J., Schurmann, C., Yaghootkar, H., D\"{o}rr, M., Frayling, T.M., Uitterlinden, A.G., Hofman, A., van Meurs, J.B, Erkeland, S.J., France, O.H. and Dehghan, A. (2014) A Genetic Variant in the Seed Region of miR-4513 Shows Pleiotropic Effects on Lipid and Glucose Homeostasis, Blood Pressure, and Coronary Artery Disease. \emph{Hum Mutat}, 35:1524–1531.

\bibitem[27]{WIN} Winham, S.J., Cuellar-Barboza, A.B., Oliveros, A., McElroy, S.L., Crow, S., Colby, C., Choi, D.S., Chauhan, M., Frye, M. and Biernacka, J.M.. (2014) Genome-wide association study of bipolar disorder accounting for effect of body mass index identifies a new risk allele in TCF7L2.\emph{Molecular Psychiatry}, 19, 1010–1016.

\bibitem[28]{BRUN} Brunham, L.R., Kruit, J.K., Verchere, C.B. and Hayden, M.R. (2008) Cholesterol in islet dysfunction and type 2 diabetes. \emph{J. Clin. Invest}, 118:403–408 (2008). doi:10.1172/JCI33296.

\bibitem[29]{MURC}Murcray, C. E., Lewinger, J. P. and Gauderman, W. J. (2009). Gene-environment interaction in genome-wide association studies. \emph{American journal of epidemiology}, 169(2), 219-226.

\bibitem[30]{THOM}Thomas, D. (2010). Methods for investigating gene-environment interactions in candidate pathway and genome-wide association studies. \emph{Annual review of public health}, 31, 21-36.

\bibitem[31]{MUKB}Mukherjee, B., Ahn, J., Gruber, S. B. and Chatterjee, N. (2012). Testing gene-environment interaction in large-scale case-control association studies: possible choices and comparisons. \emph{American journal of epidemiology}, 175(3), 177-190.

\bibitem[32]{WUCU}Wu, C. and Cui, Y. (2013). Boosting signals in gene-based association studies via efficient SNP selection. \emph{Briefings in Bioinformatics}, 15(2), 279-291.

\bibitem[33]{JZS}Jin, L., Zuo, X., Su, W., Zhao, X., Yuan, M., Han, L.,Zhao, X., Chen, Y. and Rao, S.(2014). Pathway-based analysis tools for complex diseases: a review. \emph{Genomics, Proteomics and Bioinformatics}, 12(5), 210-220.

\bibitem[34]{JHDZ}Jiang, Y., Huang, Y., Du, Y., Zhao, Y., Ren, J., Ma, S. and Wu, C. (2017). Identification of prognostic genes and pathways in lung adenocarcinoma using a Bayesian approach. \emph{Cancer Informatics}, 1(7).


\bibitem[35]{NHZ}Niu, Y. S., Hao, N., and Zhang, H. H. (2018). Interaction screening by partial correlation. \emph{Statistics and Its Interface}, 11(2), 317--325.

\bibitem[36]{XWZ}Xu, Y., Wu, M., Zhang, Q. and Ma, S. (2019). Robust identification of gene-environment interactions for prognosis using a quantile partial correlation approach. \emph{Genomics}, 111(5), 1115--1123.

\bibitem[37]{RZLM}Ren, J., Zhou, F., Li, X., Ma, S., Jiang, Y. and Wu, C. (2021). Robust Bayesian variable selection for gene-environment interactions. (under revision)

\bibitem[38]{WCM}Wu, C., Cui, Y., and Ma, S. (2014). Integrative analysis of gene–environment interactions under a multi‐response partially linear varying coefficient model. \emph{Statistics in medicine}, 33(28), 4988--4998.


\bibitem[39]{ZRLJ}Zhou, F., Ren, J., Li, G., Jiang, Y., Li, X., Wang, W. and Wu, C. (2019). Penalized Variable Selection for Lipid–Environment interactions in a longitudinal lipidomics study. \emph{Genes}, 10(12), 1002.


\bibitem[40]{MAYR}Ma, S., Yang, L., Romero, R. and Cui, Y. (2011). Varying coefficient model for gene–environment interaction: a non-linear look. \emph{Bioinformatics}, 27(15), 2119-2126.

\bibitem[41]{WUCY}Wu, C. and Cui, Y. (2013). A novel method for identifying nonlinear gene–environment interactions in case–control association studies. \emph{Human genetics}, 132(12), 1413-1425.

\bibitem[42]{ZZC}Zhao, N., Zhang, H., Clark, J. J., Maity, A., and Wu, M. C. (2019). Composite kernel machine regression based on likelihood ratio test for joint testing of genetic and gene–environment interaction effect. \emph{Biometrics}, 75(2), 625-637.

\bibitem[43]{WZCY}Wu, C., Zhong, P. S. and Cui, Y. (2018). Additive varying-coefficient model for nonlinear gene-environment interactions. \emph{Statistical applications in genetics and molecular biology}, 17(2).

\bibitem[44]{WSCM}Wu, C., Shi, X., Cui, Y. and Ma, S. (2015). A penalized robust semiparametric approach for gene–environment interactions. \emph{Statistics in medicine}, 34(30), 4016-4030.







\end{thebibliography}
\end{document}